\documentclass[aps,pra,reprint,nofootinbib,floatfix]{revtex4-1}
\usepackage{amsmath,amssymb,amsthm}
\usepackage{latexsym}
\usepackage[pdftex]{graphicx}
\usepackage{subfig}
\usepackage{color}
\usepackage[pdftex,colorlinks]{hyperref}

\newcommand{\bra}[1]{\left\langle #1 \right|}
\newcommand{\ket}[1]{\left| #1 \right\rangle}

\newcommand{\tr}{\mathrm{Tr}}
\newcommand{\gem}{\mathcal{E}} 
\newcommand{\unmax}[1]{\underline{#1}} 
\newcommand{\unc}{\unmax{C}} 
\newcommand{\Int}[0]{\int\!\!\!\!\!\int\!\!\!\!\!\int} 

\begin{document}

\title{Non-Markovian Dynamics and Entanglement of Two-level Atoms in a Common Field}

\author{C. H. Fleming}
\email{hfleming@physics.umd.edu}
\affiliation{Joint Quantum Institute, University of Maryland, College Park, Maryland 20742-4111, USA}

\author{N. I. Cummings}
\email{nickc@umd.edu}
\affiliation{Joint Quantum Institute, University of Maryland, College Park, Maryland 20742-4111, USA}

\author{Charis Anastopoulos}
\email{anastop@physics.upatras.gr}
\affiliation{Department of Physics, University of Patras, 26500 Patras, Greece}

\author{B. L. Hu}
\email{ blhu@umd.edu}
\affiliation{Joint Quantum Institute and Maryland Center for Fundamental Physics, University of Maryland, College Park, Maryland 20742-4111, USA}

\date{Jan 11, 2011}

\begin{abstract}
We derive the stochastic equations and consider the non-Markovian dynamics of a system of multiple two-level atoms in a common quantum field.
We make only the dipole approximation for the atoms and assume weak atom-field interactions.
From these assumptions we use a combination of non-secular open- and closed-system perturbation theory, and we abstain from any additional approximation schemes.
These more accurate solutions are necessary to explore several regimes: in particular, near-resonance dynamics and low-temperature behavior.
In detuned atomic systems, small variations in the system energy levels engender timescales which, in general, cannot be safely ignored, as would be the case in the rotating-wave approximation (RWA).
More problematic are the second-order solutions, which, as has been recently pointed out \cite{Accuracy},
cannot be accurately calculated using any second-order perturbative master equation, whether RWA, Born-Markov, Redfield, etc..
This latter problem, which applies to all perturbative open-system master equations, has a profound effect upon calculation of entanglement at low temperatures.
\emph{We find that even at zero temperature all initial states will undergo finite-time disentanglement} (sometimes termed ``sudden death''), in contrast to previous work.
We also use our solution, without invoking RWA, to characterize the necessary conditions for Dickie subradiance at finite temperature.
We find that the subradiant states fall into two categories at finite temperature: one that is temperature independent and one that acquires temperature dependence.
With the RWA there is no temperature dependence in any case.
\end{abstract}

\maketitle

\section{Introduction}
\subsection{Motivation}

Atomic systems constitute an important setting for the investigation of quantum decoherence and entanglement phenomena essential for quantum information processing considerations \cite{Cirac95,Raimond01,Moehring07,Ladd10}.
The physical principles underlying these systems are quite well understood, and they can be controlled and measured with great precision.
One aspect of quantum entanglement dynamics that has received significant attention is the phenomenon of entanglement sudden death, or  finite-time disentanglement, while energy and local coherences only decay away exponentially in time \cite{Zyczkowski01,YuEberly04,YuEberly09}.
A common setting for theoretical discussion of this phenomenon is atomic systems interacting with the electromagnetic field \cite{YuEberly04,Yonac06,Yonac07,Ficek06,ASH06,ASH09}, serving as an environment in the quantum open system (QOS) perspective.

When considering atomic dynamics with an open-system approach, where there is a continuum of field modes that are treated as a reservoir, the \emph{Born-Markov approximation} (BMA) \cite{Agarwal74,Carmichael99,Breuer02,QOS} is commonly invoked.
The BMA is sufficient to calculate the second-order (in the interaction Hamiltonian) dynamics of quantum open systems, even when far outside of the Markovian regime \cite{QOS}, such as low-temperature and $1/f$ noise.
However, it has been recently discovered that the second-order dynamics (more specifically, the master equation) of quantum open systems is not sufficient to fully generate the second-order solutions, including detailed state information such as entanglement \cite{Accuracy}.
The errors involved are particularly exacerbated at low temperature where the noise correlations of the environment exhibit long-ranged correlations,
whereas there are no such problems in the Markovian regime wherein the BMA is exact.
Therefore, when calculating entanglement dynamics with a second-order master equation,
careful attention must be payed to whether a positive value of the entanglement measure exceeds the inherent second-order errors.

Much of the theoretical work on atom-field systems is also derived using the \emph{rotating-wave approximation} (RWA) \cite{Agarwal74,WallsMilburn95,Carmichael99,RWA}, which eases calculation by placing the master equation into a Lindblad form \cite{Lindblad76,Gorini76,Kossakowski72,Davies74,Davies76a,Davies77,Alicki07,Accardi02,Attal06,Ingarden97,Lindblad83,Weiss93} and therefore ensuring completely-positive evolution.
In addition to weak coupling the (post-trace) RWA also requires that there be no nearly-resonant energy levels%
\footnote{Near-resonant terms can be preserved in implementing the RWA, but this will then lead to a master equation not of Lindblad-form as in \cite{Scala07}.}.
As most previous work on multipartite open systems has been performed at resonance, it is worth asking how strongly such results are dependent upon this assumption.

Until now, a systematic treatment of multiple two-level atoms in a common quantum field has yet to be carried out in a manner which can fully and accurately predict entanglement evolution and address critical issues such as sudden death of entanglement.
There are two important reasons for this.
(1) Calculations have usually involved perturbative master equations, either explicitly or by invoking the BMA or RWA which both implicitly assume a perturbative coupling to the environment%
\footnote{We would note that a non-perturbative (or higher-order) treatment of the model with an RWA system-environment interaction is still qualitatively interesting.}.
However, second-order master equations fundamentally cannot rule out second-order errors in their entanglement monotones.
This is due to the little-known fact that second-order (non-Markovian) master equations are not generally capable of providing full second-order solutions except at early times \cite{Accuracy}.
Such inescapable errors can potentially lead to entirely different and erroneous qualitative features of entanglement dynamics and sudden death.
(2) Use of the RWA also does not allow for consideration of near resonance (as additional near-stationary terms are needed in the Dirac picture)
The existence of a sub-radiant \emph{dark state} generically requires the resonance condition,
but determining how critical this is requires some analysis of the near-resonance regime.
The aim of the paper is to lead towards a complete and systematic treatment of multiple two-level atoms in a common quantum field.
To this end, the issues (1) and (2) have to be addressed and this is what we do in this paper.

\subsection{New Methods and Results}

In this work we employ the mundane second-order master equation, which we describe in Sec.~\ref{sec:SOME}, but only to explore the open-system dynamics and zeroth-order state information (e.g. dark and bright state formation).
At second order the perturbative master equation is consistent with the well-known Born-Markov and Redfield master equations, however it does not require any Markovian approximation to be derived \cite{Kampen97,Breuer03,Strunz04,QOS}.
We generally forsake the RWA so that we can explore near-resonance, however this is otherwise irrelevant in determining this timescale information\cite{RWA}.
The new method we apply in utilizing the well-known second-order master equation are the canonical perturbative solutions detailed succinctly in \cite{Accuracy} and more thoroughly in \cite{QOS}, which we also briefly outline in this paper.

To calculate the late-time entanglement of the open-system, as is vitally important to consider questions pertaining to death of entanglement,
we require second-order state information which no second-order master equation can provide \cite{Accuracy}.
In principle one could determine this information from the fourth-order master equation, which naturally excludes the BMA and RWA,
however we choose to more simply calculate the second-order asymptotic state directly from (canonical) closed system + environment perturbation theory in Sec.~\ref{sec:asymptotic}.
Note, therefore, that this second-order state information is necessarily generated by non-Markovian dynamics, even though we avoid such an open-system calculation.
We are able to calculate the second-order asymptotic state of the open system at zero temperature and at high temperatures. 
This is a new method that we have developed specifically to overcome shortcomings of the perturbative master equation discovered in \cite{Accuracy}, and more attention will be given to it in future letters.

In this way we are able to show that the two atoms in a single field are not asymptotically entangled,
even when near resonance and very close together --- which is the criterion for a dark state.
This asymptotic behavior turns out to be rather opposite to that of two oscillators in a field, which can be asymptotically entangled \cite{Lin09}.
In fact, we find that even at zero temperature the entanglement of any pair of atoms will always undergo sudden death, regardless of the initial state.
In addition to this we also find that the prediction of revival of entanglement in \cite{Ficek06} can be unreliable in some cases (due to similarly-sized errors arising from the use of the second-order master equation), unless the decay rate is sufficiently small.

Furthermore we explore the existence of sub and super-radiant (dark and bright) Dickie states at finite temperature for many atoms and with detuning for two atoms in Sec.~\ref{sec:DarkState}.
To achieve dark and bright states one requires proximity better than the resonant wavelength, as is known,
and tuning better than the ordinary dissipation rate, as we find.
The effect of finite temperature is relatively more interesting, especially when one does not employ the RWA-interaction approximation in the Hamiltonian.
There appear to be a subset of dark states, which we designate \emph{improper dark states}, which become temperature dependent.


We now briefly discuss the Markov property and its implications upon open-system dynamics.
In Sec.~\ref{sec:MicroModel} we present the microscopic model which generates our stochastic equations of motion.
In Sec.~\ref{sec:SOME} we briefly describe the perturbative second-order master equation.
Then in Sec.~\ref{sec:SOSols} we discuss the corresponding master-equation solutions, the resulting dynamics, and the accuracy of these solutions.
Finally in Sec.~\ref{sec:asymptotic}, we discuss the alternative calculation of the second-order asymptotic state,
as well as the implications upon entanglement dynamics.

\subsection{Non-Markovian Dynamics and Completely-Positive Evolution}

To clarify our language, we use the terminology \emph{Markovian} and \emph{non-Markovian} in the classical sense.
The environment and resulting dynamics of an open system are said to be non-Markovian if the underlying stochastic processes of the environment exhibit non-singular time correlations or non-vanishing correlation timescales.
In this sense, Markovian dynamics correspond to white noise whereas non-Markovian dynamics correspond to colored noise.
For a relativistic field, the noise is colored by both low-temperature (quantum) fluctuations and the (relativistic) dispersion relation (see Sec.~\ref{sec:correlations}).
This is the more physical notion of the Markov property, as given non-Markovian dynamics in this sense
(1) the quantum regression theorem does not apply \cite{Swain81},
(2) one cannot add Hamiltonian terms to the master equation post derivation (e.g. see \cite{QBM}: Sec.~7.1),
(3) one cannot add dissipative terms to the master equation post derivation (e.g. see \cite{Scala07}).

The classical usage of the terminology Markovian and non-Markovian is slightly different when applied to the master equation itself (and not its resulting dynamics).
A master equation is said to have a \emph{non-Markovian representation} when placed into integro-differential form, e.g.
\begin{align}
\dot{\boldsymbol{\rho}}(t) &= \int_0^t \!\! d\tau \, \boldsymbol{\mathcal{K}}(t \!-\! \tau) \, \boldsymbol{\rho}(\tau) \, ,
\end{align}
whereas the master equation is said to have a \emph{Markovian representation} when placed into a time-local form, e.g.
\begin{align}
\dot{\boldsymbol{\rho}}(t) &= \boldsymbol{\mathcal{L}}(t) \, \boldsymbol{\rho}(t) \, .
\end{align}
Obviously this distinction is superficial here as one can trivially relate these two master equations
\begin{align}
\boldsymbol{\mathcal{L}}(t) &= \dot{\boldsymbol{\mathcal{G}}}(t) \, \boldsymbol{\mathcal{G}}(t)^{-1} \, , \\
\hat{\boldsymbol{\mathcal{G}}}(s) &= \left[ s - \hat{\boldsymbol{\mathcal{K}}}(s) \right]^{-1} \, ,
\end{align}
in terms of the Laplace transform, defined $\hat{f}(s) = \int_0^\infty \! dt \, e^{-st} \, f(t)$.
While Markovian dynamics imply a Markovian representation, the inverse is not always true, e.g. see the Hu-Paz-Zhang master equation \cite{HPZ92,QBM}.
Time-dependence in $\boldsymbol{\mathcal{L}}(t)$ is also not specific to non-Markovian dynamics,
as one can couple a time-dependent Hamiltonian system to a Markovian environment.

These distinctions are further complicated for Lindblad master equations obtained after the post-trace RWA \cite{RWA}.
Given their mathematical structure, such master equations can be unraveled with Markovian processes,
even if the microscopic environment from which they were derived was not Markovian (or even close to being Markovian).
This Markovian theory is essentially an effective theory, which is sufficient for modeling certain aspects of the perturbative dynamics. 

One might question how positivity can be maintained with a master equation which is not of Lindblad form.
In general, for non-Markovian dynamics, master equations need not be of Linblad form to ensure completely-positive evolution.
Indeed, there are exact master equations known which do not have Lindblad form and yet have a time-local or Markovian representation, e.g. see the Hu-Paz-Zhang master equation \cite{HPZ92,QBM}.
The positivity, or lack thereof, generated by a perturbative master equation not of Lindblad form can be understood as follows.
If we can produce solutions $\boldsymbol{\rho}(t)$ which are good to some order $\mathcal{O}(g^{2n})$ in the system-environment coupling, parameterized by $g$,
and the higher-order errors are nonsecular in time,
then positivity is only ensured to that order, or more specifically
\begin{align}
\boldsymbol{\psi}^\dagger \boldsymbol{\rho}(t) \, \boldsymbol{\psi} &> 0 + \mathcal{O}(g^{2n+2}) \, , \label{eq:ppos}
\end{align}
for all Hilbert-space vectors $\boldsymbol{\psi}$.
Manipulating the master equation into a Lindblad form will produce solutions which satisfy Eq.~\eqref{eq:ppos} exactly,
however they are not necessarily more accurate in any other sense.
The pseudo-Lindblad master equation will preserve the trace and Hermiticity of the density matrix exactly,
and the master equation will belong to the appropriate class of algebraic generators (being the difference of two Lindblad generators).

\section{The Microscopic Model}\label{sec:MicroModel}

\subsection{Interacting Hamiltonians}
We wish to investigate the properties of multiple atoms interacting with a common electromagnetic field in free space, which serves as the environment in the open quantum system description.
We will consider the atoms to be held stationary at fixed positions $\mathbf{r}_n$, and only consider the dynamics of the electrons in orbit.
As we consider neutral atoms far apart, we will neglect the electrostatic interactions between the atoms.
We will use the two-level approximation to describe the atoms, so that they are an array of, otherwise non-interacting, qubits.
Only the lowest-level excitations of the Dirac field are considered (see for instance, \cite{Anastopoulos00}, App. A), so that we have
\begin{align}
\mathbf{H}_\mathrm{sys + field} &= \mathbf{H}_\mathrm{sys} + \mathbf{H}_\mathrm{int} + \mathbf{H}_\mathrm{field} \, , \label{eq:H2LA} \\
\mathbf{H}_\mathrm{sys} &= \sum_n \Omega_n \, \boldsymbol{\sigma}_{\!+_n} \boldsymbol{\sigma}_{\!-_n} \, , \\
\mathbf{H}_\mathrm{int} &= \sum_n \boldsymbol{\sigma}_{\!x_n} \, \mathbf{d}_n \cdot \mathbf{A}(\mathbf{r}_n) \, ,
\end{align}
where $\boldsymbol{\sigma}$ denote the Pauli matrices and $\mathbf{d}$ corresponds to the lowest-energy levels of the electron's dipole matrix \cite{Agarwal71,WallsMilburn95,Anastopoulos00}.
The field Hamiltonian and vector potential are given by
\begin{align}
\mathbf{H}_\mathrm{field} &= \Int \!\! d^3k \sum_{\boldsymbol{\epsilon}_k} \omega_k \, \mathbf{a}^\dagger_{\mathbf{k},\boldsymbol{\epsilon}_k} \mathbf{a}_{\mathbf{k},\boldsymbol{\epsilon}_k} \, , \\
\mathbf{A}(\mathbf{x}) &= \frac{1}{(2\pi)^{3/2}} \Int \!\! d^3k \sum_{\boldsymbol{\epsilon}_k} \mathbf{A}_{\mathbf{k},\boldsymbol{\epsilon}_k}(\mathbf{x}) \, , \label{eq:Ak} \\
\mathbf{A}_{\mathbf{k},\boldsymbol{\epsilon}_k}(\mathbf{x}) &= \frac{\boldsymbol{\epsilon}_k}{\sqrt{2\varepsilon_0\omega_k}} \left\{ e^{+\imath \mathbf{k} \cdot \mathbf{x}} \, \mathbf{a}_{\mathbf{k},\boldsymbol{\epsilon}_k} + e^{-\imath \mathbf{k} \cdot \mathbf{x}} \, \mathbf{a}_{\mathbf{k},\boldsymbol{\epsilon}_k}^\dagger \right\} \, ,
\end{align}
with $k = \omega_k / c$ and where $\boldsymbol{\epsilon}_{\mathbf{k}}$ denote the polarization vectors perpendicular to $\mathbf{k}$ \cite{Milonni93,CohenTannoudji97}.
We have assumed the atoms to be relatively stationary and that the atomic transition in each atom will produce linearly polarized photons (i.e., both ground and excited state are eigenstates of some component of angular momentum with the same eigenvalue).


Even though we assume the atoms to be ideal dipoles, we do neglect the $1/r^3$ electrostatic interactions between the electronic dipole moments,
and treat $r$ as a classical parameter rather than a quantum degree of freedom.
Thus we necessarily assume a sufficient degree of separation between all atoms.
It will be seen that magnetostatic interactions between the atomic dipole moments will be introduced via backreaction from the field,
but these will be terms will be $1/r$ and thus can play a role even in the regime where the electrostatic interactions are negligible.

\subsection{Environment Correlations}\label{sec:correlations}
Any perturbative open-system analysis of this problem, whether the master equation we employ here or Langevin equation, 
will entail calculation of the correlation function for the field we couple to
\begin{equation}
\boldsymbol{\alpha}_{nm}(t,\tau) = \left\langle \mathbf{d}_n^\mathrm{T} \, \underline{\mathbf{A}}(\mathbf{r}_n;t) \, \underline{\mathbf{A}}(\mathbf{r}_m;\tau)^\mathrm{T} \, \mathbf{d}_m \right\rangle_{\!\mathrm{field}} \, ,
\end{equation}
where $\underline{\mathbf{A}}(\mathbf{r};t)$ denotes the time-dependent vector potential $\mathbf{A}(\mathbf{r})$ in the interaction or Dirac picture.
All perturbative master-equation coefficients in Sec.~\ref{sec:SOME} will be generated by these free-field correlations.
In addition to being Hermitian and positive definite, for our consideration of stationary atoms, the correlation function is also stationary: $\boldsymbol{\alpha}_{nm}(t,\tau) = \boldsymbol{\alpha}_{nm}(t \!-\! \tau)$.
For simplicity we will further assume that all system dipole moments are of the same magnitude in this work, though such differences can be easily accounted for.

The fluctuation-dissipation relation (FDR) \cite{FDR} allows us to express the environmental correlations in terms of the damping kernel as
\begin{align}
\tilde{\boldsymbol{\alpha}}(\omega) &= \tilde{\boldsymbol{\gamma}}(\omega) \frac{\omega}{\sinh\!\left(\frac{\omega}{2T}\right)} e^{-\frac{\omega}{2T}} \, , \label{eq:alphaThermal} \\
& = 2 \, \tilde{\boldsymbol{\gamma}}(\omega) \, \omega  \left[ \bar{n}\! \left( \left| \omega \right|, T \right) + \mathrm{\theta}\! \left( -\omega \right) \right] \, , 
\end{align}
here in the Fourier domain, defined
\begin{equation}
\tilde{f}(\omega) \equiv \int_{-\infty}^{+\infty} \!\!\! dt \, e^{-\imath \omega t} \, f(t)
\end{equation}
and where $ \bar{n}\! \left(\omega, T \right) $ is the thermal average photon number in a mode of frequency $ \omega $.

The damping kernel is formed by commutators of the vector potential and therefore in these models it is independent of the state of the environment (temperature independent)
and also the same whether in the classical or quantum regime.
Outside of the high-temperature (semi-classical) regime where $\tilde{\boldsymbol{\alpha}}(\omega) = 2 \, T \, \tilde{\boldsymbol{\gamma}}(\omega)$,
the damping kernel is the simpler object to inspect,
and in the free-particle Hamiltonian corresponding to \eqref{eq:H2LA} it directly produces the Abraham-Lorentz force \cite{ADL}.
At least peturbatively, the damping kernel is only responsible for irreversible dissipation of energy, whereas the full correlation also contains the influence of noise \cite{FDR}.

\subsubsection{Scalar-Field Correlations}\label{sec:gammaScalar}
First let us consider the simpler case of a scalar field, wherein one essentially neglects the polarization vectors in $\mathbf{A}$ and the direction of atomic dipole moments.
In this case we calculate the associated damping kernel to be
\begin{align}
\tilde{\gamma}_{nm}(\omega) &= \tilde{\gamma}_{0} \, \mathrm{sinc}(r_{nm} \omega) \, , \label{eq:gammaScalar}
\end{align}
where $\mathbf{r}_{nm} = \mathbf{r}_n \!-\! \mathbf{r}_m$ is the difference vector for the positions of the two atoms
and $\tilde{\gamma}_{0} = \tilde{\gamma}_{nn}(\omega)$ is the self-damping coefficient.
Restoring factors of $c$, the sinc function appears to act as a relativistic regulator for $\omega \ll c/r$.
Therefore, in the nonrelativistic regime one only has local damping for charges at precisely the same location.
In fact, this must always be the case for linear coupling to a relativistic field with a position-like field operator such as $\mathbf{A}$.
The damping kernel is deterministic and the same whether in the classical or quantum regimes.
Thus it is necessarily constrained to the light cone.

Note for instance the temporal representation of the scalar-field damping kernel
\begin{align}
\gamma_{nm}(t) &= \frac{\tilde{\gamma}_{0}}{2} \delta_{r_{nm}}\!(t) \, , \label{eq:g(t)} \\
\delta_r(t) &\equiv \frac{\mathrm{\theta}(r\!-\!|t|)}{2 r} \, ,
\end{align}
where $ \mathrm{\theta} $ is the Heaviside step function.
This kernel strictly adheres to the light cone.

\subsubsection{Electromagnetic-Field Correlations}\label{sec:EMgamma}
The electromagnetic damping kernel is slightly more complicated as it contains orientation dependence from the polarization vectors.
We calculate this kernel to be
\begin{align}
\tilde{\boldsymbol{\gamma}}_{nm}(\omega) &= \tilde{\gamma}_{0} \, \mathbf{d}_n^\mathrm{T} \left\{ \mathrm{FS}_1(r_{nm} \omega) + \mathrm{FS}_0(r_{nm} \omega) \, \hat{\mathbf{r}}_{nm} \, \hat{\mathbf{r}}_{nm}^\mathrm{T} \right\} \mathbf{d}_m \, ,  \label{eq:EMgamma} \\
\tilde{\gamma}_{0} &= \frac{1}{6 \pi \varepsilon_0 c^3} \, ,
\end{align}
in the Fourier domain.
Instead of a sinc functions, we have the related entire functions
\begin{align}
\mathrm{FS}_1(z) &\equiv +\frac{3}{2} \frac{(z^2-1)\sin(z) + z \cos(z)}{z^3} \, , \\
\mathrm{FS}_0(z) &\equiv -\frac{3}{2} \frac{(z^2-3)\sin(z) + 3 \, z \cos(z)}{z^3} \, .
\end{align}
\begin{figure}
\includegraphics[width=0.4\textwidth]{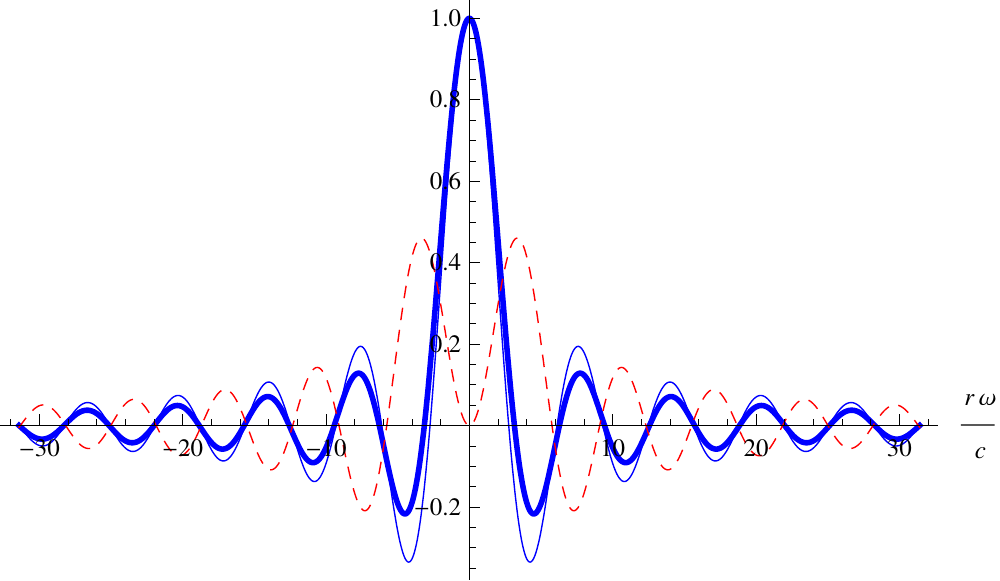}
\caption{Comparison of sinc (bold), $\mathrm{FS}_1$, and $\mathrm{FS}_0$ (dashed). Sinc and $\mathrm{FS}_1$ are extremely qualitatively similar, both being unity at zero whereas $\mathrm{FS}_0$ vanishes at zero.}
\label{fig:Polarity}
\end{figure}
\begin{figure}
\includegraphics[width=0.4\textwidth]{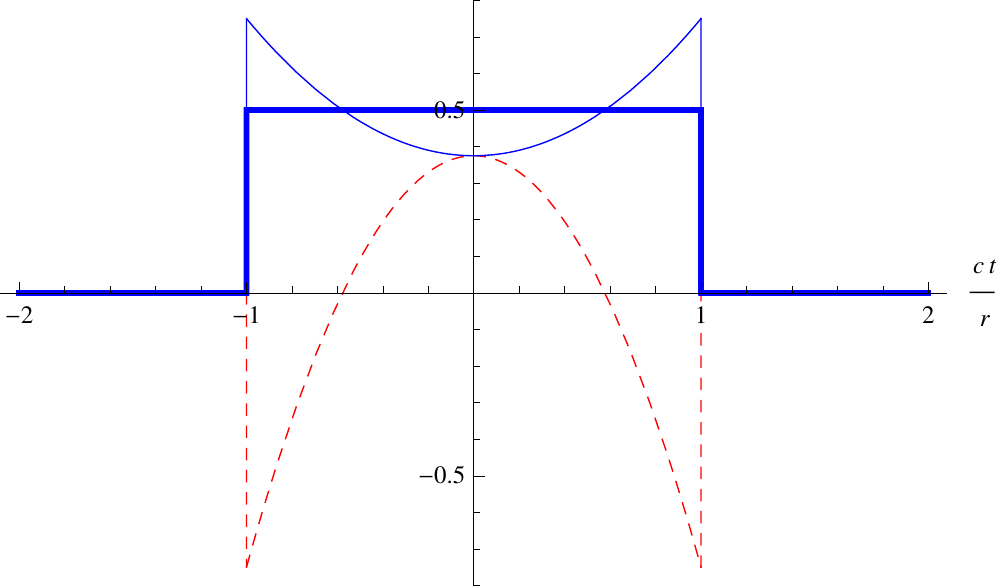}
\caption{Comparison of the same functions in Fig.~\ref{fig:Polarity}, but in the time domain: the rectilinear distribution (bold), $\mathrm{FS}_1$, and $\mathrm{FS}_0$ (dashed).
All of these functions are constrained to the light cone.}
\label{fig:Polarity2}
\end{figure}
In Fig.~\ref{fig:Polarity} we compare these functions.
In any case, the cross correlations are very nonlocal and are maximized when the atoms are close.
Whereas when the atoms are very far apart, the cross correlations always vanish and thus all noise can be treated independently.
One can see that the scalar-field correlations are very similar to that of the electromagnetic field when $\mathbf{d}_n \parallel \mathbf{d}_m \perp \mathbf{r}_{nm}$.
As we will wish to maximize cross correlations, we will primarily work with the scalar-field correlations, which are very similar to the problem of parallel dipoles in the electromagnetic field.

\section{Second-order master equation}\label{sec:SOME}
\subsection{Master equation and coefficients}\label{sec:coefficients}
The second-order master equation for the reduced density matrix of the dipoles can be expressed \cite{QOS}
\begin{align}
\dot{\boldsymbol{\rho}} &= \boldsymbol{\mathcal{L}} \, \boldsymbol{\rho} \, , \\
\boldsymbol{\mathcal{L}} &= \boldsymbol{\mathcal{L}}_0  + \boldsymbol{\mathcal{L}}_2 + \cdots
\end{align}
in terms of the zeroth and second-order Liouville operators
\begin{align}
\boldsymbol{\mathcal{L}}_0 \, \boldsymbol{\rho} &=  \left[ -\imath \mathbf{H}, \boldsymbol{\rho} \right] \, , \\
\boldsymbol{\mathcal{L}}_2 \, \boldsymbol{\rho} &= \sum_{nm} \left[ \boldsymbol{\sigma}_{\!x_n}, \boldsymbol{\rho} \, (\boldsymbol{\mathcal{A}}_{nm}\! \diamond \boldsymbol{\sigma}_{\!x_m})^\dagger - (\boldsymbol{\mathcal{A}}_{nm}\! \diamond \boldsymbol{\sigma}_{\!x_m}) \, \boldsymbol{\rho} \right] \, . \label{eq:WCGME}
\end{align}
We present the stationary limit of this master equation for the scalar-field environment,
by way of \eqref{eq:Hadamard1}, \eqref{eq:A(omega)}, \eqref{eq:alphaThermal}, \eqref{eq:gammaScalar}, \eqref{eq:ImAFT} (finite temperature) and \eqref{eq:ImAPT} (low-temperature expansion).
Some aspects of the electromagnetic-field environment and full-time theory are also discussed herein.
We choose to present more results for the scalar-field environment merely as a means of presenting simpler expressions which retain the qualitative features we are interested in.
The electromagnetic theory can be resolved without a considerable amount of additional difficulty, but the resulting expressions will be several times more lengthy.

The second-order operators in master equation \eqref{eq:WCGME} are most easily represented by the ladder operators as
\begin{align}
(\boldsymbol{\mathcal{A}}_{nm}\! \diamond \boldsymbol{\sigma}_{\!x_m}) &= \mathcal{A}_{nm}\!\left( +\Omega_m \right) \, \boldsymbol{\sigma}_{\!+_m} + \mathcal{A}_{nm}\!\left( -\Omega_m \right) \, \boldsymbol{\sigma}_{\!-_m}  \, , \label{eq:Hadamard1} \\
\boldsymbol{\sigma}_{\!\pm} & \equiv  \frac{1}{2} \left[ \boldsymbol{\sigma}_{\!x} + \imath \, \boldsymbol{\sigma}_{\!y} \right] \, ,
\end{align}
and the second-order coefficients being generated from the field correlations as
\begin{align}
\mathcal{A}_{nm}\!\left( \omega \right) &= \frac{1}{2} \tilde{\alpha}_{nm}(\omega) - \imath \, \mathcal{P}\!\left[ \frac{1}{\omega} \right] \! * \tilde{\alpha}_{nm}(\omega) \, , \label{eq:A(omega)}
\end{align}
here in the late-time limit (as compared to system and cutoff timescales), where $\mathcal{P}$ denotes the Cauchy principal value.
Higher-order master equation coefficients will entail convolutions over several copies of the field correlations combined with several products of the system coupling operator.

The first portion of the second-order coefficient, or Hermitian part (here real), is immediately given by Eq.~\eqref{eq:alphaThermal}.
Whereas the second term, or anti-Hermitian part (here imaginary), must be evaluated via the convolution
\begin{align}
\mathrm{Im}\!\left[ \mathcal{A}_{nm}(\omega) \right] &= -\frac{1}{2\pi} \int_{\!-\infty}^{+\infty} \!\!\! d\varepsilon \, \mathcal{P}\!\left[ \frac{ 1 }{\omega - \varepsilon} \right] \tilde{\alpha}_{nm}(\varepsilon) \, ,
\end{align}
and together they form a causal response function.
These are the coefficients which often require regularization and renormalization.
For now let us simply evaluate the bare coefficients for non-vanishing $r$.
For finite temperatures, the coefficients exactly evaluate to
\begin{align}
& \mathrm{Im}\!\left[ \mathcal{A}_{nm}(\omega) \right] = \label{eq:ImAFT}
 + \frac{\tilde{\gamma}_{0}}{r_{\!nm}} \frac{1}{\pi} \mathrm{Im}\!\left[ \Phi_1\!\left( \frac{\imath\,\omega}{2\pi T} ; 2 \pi T r_{\!nm} \right) \right] \nonumber \\
& -\frac{\tilde{\gamma}_{0}}{r_{\!nm}} \left\{ \frac{T}{\omega} - \frac{1}{2} \left[ \coth\!\left(\frac{\omega}{2T}\right)- 1 \right] \cos(r_{\!nm} \omega) \right\} \, ,
\end{align}
in terms of the Lerch $\Phi_1$ function
\begin{eqnarray}
\Phi_1(z;\lambda) &\equiv& \sum_{k=1}^\infty \frac{e^{-k \lambda}}{k + z} \, .
\end{eqnarray}
This functional representation is exact, though best for positive temperature.
Conversely, one also has the low-temperature expansion
\begin{align}
& \mathrm{Im}\!\left[ \mathcal{A}_{nm}(\omega) \right] = \frac{\tilde{\gamma}_{0}}{r_{\!nm}} \frac{\mathrm{sign}(\omega)}{\pi} \sum_{k=1}^\infty \mathrm{S}_k \label{eq:ImAPT} \\
& -\frac{\tilde{\gamma}_{0}}{r_{\!nm}} \frac{1}{\pi} \left[ \sin( r_{\!nm} \omega ) \, \mathrm{ci}( |r_{\!nm} \omega| ) - \cos( r_{\!nm} \omega ) \, \mathrm{si}( r_{\!nm} \omega ) \right] \, , \nonumber
\end{align}
in terms of the summand
\begin{align}
\mathrm{S}_k &= \frac{\mathrm{Ei}[(+k \beta + \imath \, r_{\!nm})|\omega|]}{e^{(+k \beta + \imath \, r_{\!nm})|\omega|}} + \frac{\mathrm{Ei}[(-k \beta + \imath \, r_{\!nm})|\omega|] - \imath \, \pi}{e^{(-k \beta + \imath \, r_{\!nm})|\omega|}} \, ,
\end{align}
and where the trigonometric integrals are defined
\begin{eqnarray}
\mathrm{si}(z) &\equiv& -\int_z^\infty \!\! dz' \frac{\sin(z')}{z'} \, , \\
\mathrm{ci}(z) &\equiv& -\int_z^\infty \!\! dz' \frac{\cos(z')}{z'} \, , \\
\mathrm{Ei}(z) & \equiv & -\int_{-z}^\infty \!\! dz' \mathcal{P}\!\left[ \frac{e^{-z'}}{z'} \right] \, ,
\end{eqnarray}
however, for positive temperatures this expansion is not well behaved for small energy differences.
For zero temperature, the exact relation (the second line in \eqref{eq:ImAPT})  is well behaved and matches perfectly to the zero-temperature limit of Eq.~\eqref{eq:ImAFT}.

\subsection{Unitary Dynamics and Renormalization}\label{sec:renormalization}
First we would like to focus upon the unitary generators that have been induced by the environment.
If we express the master equation in the standard pseudo-Lindblad form
\begin{align}
\boldsymbol{\mathcal{L}}_2 \! \left\{ \boldsymbol{\rho} \right\} &= \left[ -\imath \mathbf{U}, \boldsymbol{\rho} \right] + \boldsymbol{\mathcal{D}}\!\left\{ \boldsymbol{\rho} \right\} \, ,
\end{align}
then the unitary generator $\mathbf{U}$ is given by
\begin{align}
\mathbf{U} &=  \frac{1}{2\imath} \sum_{nm} \left[ \boldsymbol{\sigma}_{\!x_n} \, (\boldsymbol{\mathcal{A}}_{nm}\! \diamond \boldsymbol{\sigma}_{\!x_m}) - (\boldsymbol{\mathcal{A}}_{nm}\! \diamond \boldsymbol{\sigma}_{\!x_m})^\dagger \, \boldsymbol{\sigma}_{\!x_n} \right] \, .
\end{align}
One must be careful not to identify all of $\mathbf{U}$ with environmentally-induced forces, such as dissipation and renormalization.
The delineation between homogeneous backreaction and noise is not as clear in the master equation as it is in the Langevin equation (the Heisenberg equations of motion for the system operators as driven by the field).
The Langevin equation instructs us that only the self interactions generated by $\mathrm{Im}[\mathcal{A}_{nn}(0)] = -\gamma_{nn}(0)$ should be renormalized \cite{QOS,ADL}.

From the master-equation perspective this prescription can also be motivated, though it is more difficult to immediately see.
The self terms $\mathbf{U}_{\!nn}$ happen to be the most divergent.
These terms resolve to be
\begin{align}
\mathbf{U}_{\!nn} &= \mathrm{Im}[\mathcal{A}_{nn}(+\Omega_n)] \, \boldsymbol{\sigma}_{\!-_n} \, \boldsymbol{\sigma}_{\!+_n} + \mathrm{Im}[\mathcal{A}_{nn}(-\Omega_n)] \, \boldsymbol{\sigma}_{\!+_n} \, \boldsymbol{\sigma}_{\!-_n} \, .
\end{align}
The most divergent part of this expression is not present in the energy-level separation between the ground and excited states,
but only in the average or absolute of the energy levels.
Therefore if we choose the renormalization suggested by the Langevin equation
\begin{align}
\mathbf{U}_\mathrm{ren} &= \sum_n \boldsymbol{\sigma}_{\!x_n} \, \mathrm{Im}[\mathcal{A}_{nn}(0)] \, \boldsymbol{\sigma}_{\!x_n} \, ,
\end{align}
then the divergent contribution from
\begin{eqnarray}
\mathrm{Im}\!\left[ \mathcal{A}_{nn}(0) \right] &=& -\frac{\tilde{\gamma}_{0}}{2\, r_{\!nn}} \, , \label{eq:Ren}
\end{eqnarray}
will indeed be removed.

It is also physically instructive to inspect the corresponding cross terms from $\mathrm{Im}[\mathcal{A}_{nm}(0)]$,
as the Langevin equation reveals these terms to be non-dissipative backreaction mediated by the field (with $\mathrm{Im}[\mathcal{A}_{nn}(0)]$ being their corresponding divergent self interaction).
For the electromagnetic field these terms can be assigned the interaction potential
\begin{align}
\mathbf{V}_\mathrm{2LA} &= -\frac{\mu_0}{8 \pi}\sum_{n<m} \boldsymbol{\sigma}_{\!x_n} \, \mathbf{d}_n^\mathrm{T} \frac{ 1 + \hat{\mathbf{r}}_{nm} \, \hat{\mathbf{r}}_{nm}^\mathrm{T} }{r_{nm}} \mathbf{d}_m \, \boldsymbol{\sigma}_{\!x_m} \, ,
\end{align}
which is qualitatively approximated by the scalar field interaction for $\mathbf{d}_n \parallel \mathbf{d}_m \perp \mathbf{r}_{nm}$.
This result for the two-level atoms should be compared to the well-known magnetostatic potential of the (classical) Darwin Hamiltonian for point particles
\begin{align}
\mathbf{V}_\mathrm{Darwin} &= -\frac{\mu_0}{8 \pi}\sum_{n<m} \left[ \frac{e_n}{m_n} \mathbf{p}_n \right]^\mathrm{\!T} \frac{ 1 + \hat{\mathbf{r}}_{nm} \, \hat{\mathbf{r}}_{nm}^\mathrm{T} }{r_{nm}} \left[ \frac{e_m}{m_m} \mathbf{p}_m \right] \, ,
\end{align}
and so these two-level atoms experience magnetostatic attraction and repulsion depending upon their relative momentum states,
just as moving point charges do.

\subsubsection{Regularization}
Although we have renormalized away the $1/r_{nn}$ divergences present in Im$[A_{nn}(0)]$,
there remain logarithmic $r_{nn}$ divergences in Im$[A_{nn}(\Omega_n)]$ which cannot be renormalized,
as they extend into the dissipative portion of the master equation.
These divergences must be made finite with regularization.

Note that sinc$(\omega/\Lambda)$ is a high frequency regulator:
sinc$(z): [0,\infty) \to [1,0)$ sufficiently fast for all of our integrals to converge.
Therefore we don't need to consider any additional regularization in our damping kernel if we do not evaluate sinc$(r \omega)$ for vanishing $r$.
Instead of allowing $r$ to vanish for self-correlations, we impose a high frequency cutoff $r_0 = \Lambda^{-1}$,
perhaps motivated by the non-vanishing physical size of the dipole.
The more common alternative is to introduce cutoff regularization directly into the field coupling $\mathbf{A}_k$ in Eq.~\eqref{eq:Ak},
often by treating the coupling strength as a form factor with some gradual $\mathbf{k}$-dependence.
Different choices of cutoff regulators will yield the same results to highest order in $\Lambda$, and the theory should be somewhat insensitive to these details in the end.

\subsection{Full-time non-Markovian dynamics}
For the full-time evolution of initially uncorrelated states,
one must apply the full-time coefficients (again we work from the notation of Ref.~\cite{QOS})
\begin{align}
\mathcal{A}_{nm}(\omega;t) &= \int_0^t \!\! d\tau \, e^{-\imath \omega \tau} \, \alpha_{nm}(\tau) \, ,
\end{align}
which exhibits causal behavior in as much as the field correlations are causal.
In the semiclassical regime the field correlations are perfectly causal as $\boldsymbol{\alpha} = 2 T \boldsymbol{\gamma}$,
and the damping kernel $\boldsymbol{\gamma}$ is classical and perfectly causal (see Fig.~\ref{fig:Polarity2}).
However, in the quantum regime low-temperature fluctuations are allowed to leak outside of the light cone.

\begin{figure*}[htb]
\includegraphics[width=0.9\textwidth]{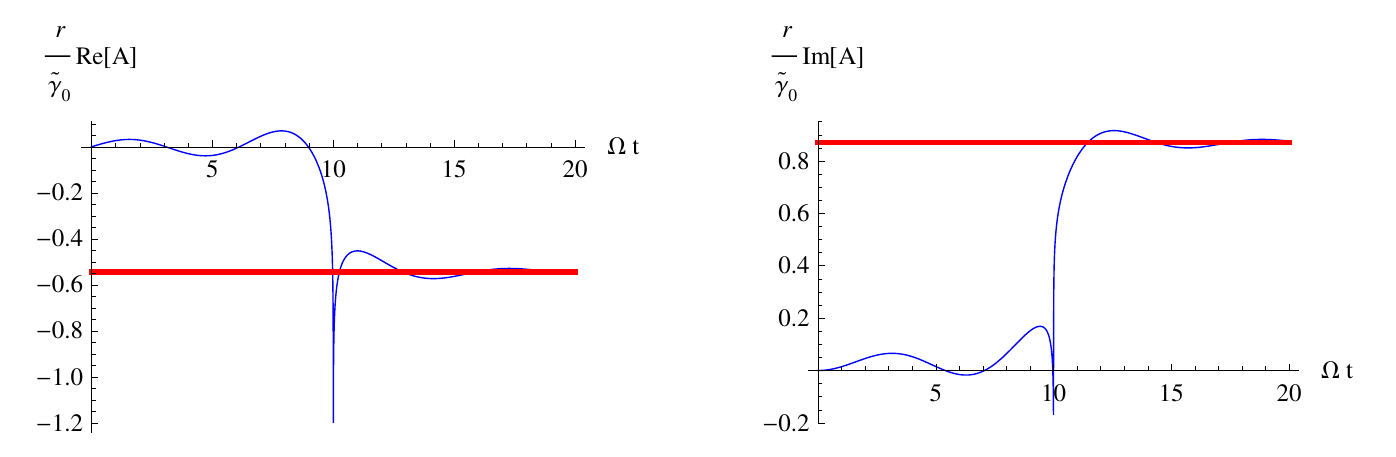}
\caption{Re$[\mathcal{A}_{nm}(-\Omega_m;t)]$ (left) and Im$[\mathcal{A}_{nm}(-\Omega_m;t)]$ (right) for a zero temperature reservoir at $r_{nm} = 10 / \Omega_m$.
The bold line denotes the asymptotic coefficients.}
\label{fig:A(t)}
\end{figure*}

If we start with factorized initial conditions for our atoms and field, then classically they would have no knowledge of each other's existence until the mediating photons could reach them at $t=r/c$.
Thus the dynamics would be perfectly factorized until this time (a highly non-Markovian behavior).
In the quantum regime, the story is slightly different as unavoidable quantum fluctuations do leak from one atom to the other, in a manner which only roughly adheres to the light cone.
We plot the resulting dynamics of the coefficients in Fig.~\ref{fig:A(t)}, where one can see the master-equation coefficients jolt at precisely $t=r/c$.
Prior to the jolt, the master equation coefficients (for finite $r$) are roughly zero and the dynamics factorized;
whereas after the jolt, the coefficients are roughly their asymptotic value and dynamics interrelated.

Shortly after $t=r/c$, one has what is essentially a joint state of the system which is properly correlated with the field.
Properly-correlated initial states can be more accurately modeled by a more laboratory-appropriate preparation scheme wherein a desired state of the open system is properly prepared from a global equilibrium state,
yet without significantly disturbing the vital system-environment equilibrium correlations \cite{Correlations}.
However in the weak-coupling regime it is sufficient to consider the asymptotic dynamics.
Therefore from hereon we will consider $\boldsymbol{\mathcal{L}}(\infty)$ as approximately corresponding to the open-system dynamics of properly-correlated initial states,
and not the highly non-Markovian dynamics of factorized initial states.

\section{Second-order dynamics}\label{sec:SOSols}
\subsection{Canonical perturbation theory}
The open-system dynamics of properly-correlated initial states are approximately described by the time-independent Liouvillian
$\boldsymbol{\mathcal{L}}(\infty)$, which we will now write simply as $ \boldsymbol{\mathcal{L}} $.
The time evolution is then approximately
$ e^{t \, \boldsymbol{\mathcal{L}}} $, and it can be computed (analogously to the time-independent Schr\"{o}dinger equation) simply from the solutions of the eigen-value problem
\begin{eqnarray}
\boldsymbol{\mathcal{L}} \{ \mathbf{o} \} &=& f \, \mathbf{o} \, ,
\end{eqnarray}
where $f$ is an eigen-frequency and $\mathbf{o}$ a right eigen-operator (super-vector)%
\footnote{These eigen-operators are often referred to as the damping basis of the master equation\cite{Briegel93}.}.
In principle this can be performed numerically with the super-matrix operators, but to avoid more costly numerics we have resorted to a careful application of canonical perturbation theory, as can found in Ref.~\cite{QOS}.
Because the master equation itself is perturbative, there is no loss in accuracy by finding the solutions perturbatively.

It is of paramount importance to note that second-order master equations can only determine all eigen-values $f$ to second order \cite{Accuracy}.
This includes the coherent oscillation, emission, and decoherence rates.
However, entanglement and other state information is obtained from the eigen-operators $\mathbf{o}$, which second-order master equations can generally only determine to zeroth order \cite{Accuracy}.
Therefore in this section we only address dynamical questions answered by $f$, and in Sec.~\ref{sec:asymptotic} we will calculate and inspect $\boldsymbol{\rho}(\infty)$ to second order,
by using an alternative mathematical formalism which has similar ingredients.

At zeroth-order our eigen-value problem corresponds to the energy-level differences and outer-products of energy states
\begin{eqnarray}
\boldsymbol{\mathcal{L}}_0 \{ \ket{\omega_i}\!\!\bra{\omega_j} \} &=& -\imath \, \omega_{ij} \ket{\omega_i}\!\!\bra{\omega_j} \, ,
\end{eqnarray}
where $\omega_{ij} = \omega_{i} - \omega_{j}$.
The environment induces frequency shifts (including decay) and basis corrections such that the eigen-operators are no longer dyadic in any basis of Hilbert-space vectors.
Some degree of degeneracy is also inescapable as $\omega_{ii} = \omega_{jj}=0$.

As our system coupling is non-stationary, with no additional degeneracies the cross-coupling will have no effect upon the second-order frequencies of the perturbed off-diagonal operators,
and the $f_{ij}$ corresponding to $\ket{\omega_i}\!\!\bra{\omega_j}$ for $i \neq j$ are given by
\begin{align}
f_{ij} &= -\imath \, \omega_{ij} + \bra{\omega_i} \boldsymbol{\mathcal{L}}_2 \! \left\{ \ket{\omega_i}\!\!\bra{\omega_j} \right\} \ket{\omega_j} \, ,
\end{align}
which reference no cross-correlations.
Second-order corrections to the eigen-operators $\mathbf{o}$ (and thus states)  can then be found by perturbative consistency with the master equation.
Dynamics of the diagonal operators and any other degenerate (and near-degenerate) subspaces must be treated much more carefully with degenerate perturbation theory.
For the energy states, their second-order dynamics are encapsulated by a Pauli master equation.
This gives rise to their second-order relaxation rates and zeroth-order eigen-operators.
Due to inherent degeneracy, $\omega_{ii}=\omega_{jj}=0$ and any resonant frequencies, their second-order operator perturbations require the fourth-order Pauli master equation \cite{Accuracy,QOS}.
To summarize, in general the matrix elements of the solution $ \boldsymbol{\rho}(t) $ expressed in the (free) energy basis will be accurate to $ \mathcal{O}(\gamma) $ off the diagonal but only to $\mathcal{O}(\gamma^0)$ on the diagonal (though timescales are known to $ \mathcal{O}(\gamma) $).  This inaccuracy in the diagonals is an inherent limitation of any perturbative master equation, including those derived under the RWA or the BMA \cite{Accuracy}. With the RWA, however, \emph{all} matrix elements are only good to $\mathcal{O}(\gamma^0)$%
\footnote{When looking only at observables time-averaged over many system periods $ 2 \pi / \Omega $ some of these additional discrepancies generated by the RWA can be greatly reduced.}.

\begin{figure}
\includegraphics[width=0.4\textwidth]{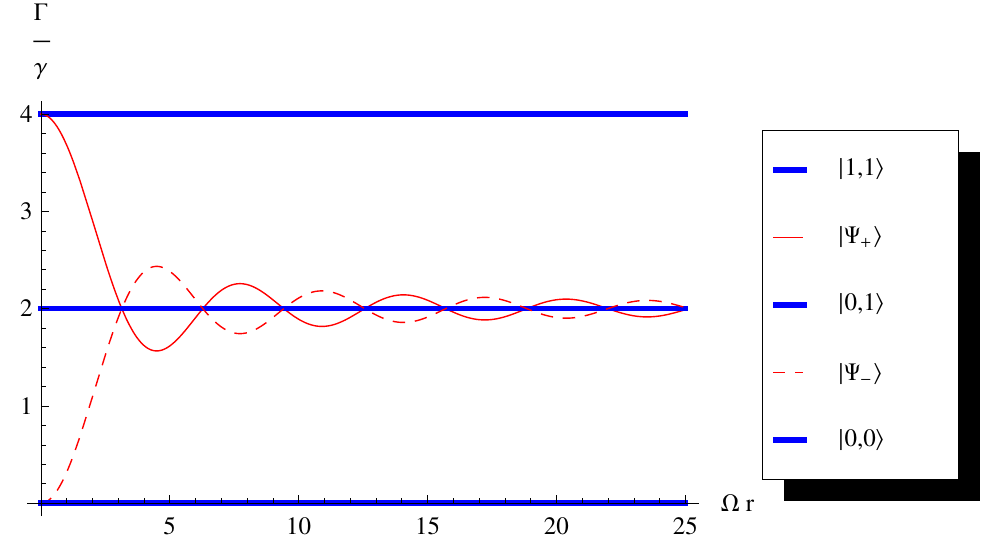}
\caption{Decay rates of the (zeroth-order) stationary operators for two resonant atoms in a zero-temperature environment at varying separation distance.
The legend indicates the pure states they approximately correspond to in the order they occur at the vertical axis.}
\label{fig:Decay1a}
\end{figure}
\begin{figure}
\includegraphics[width=0.4\textwidth]{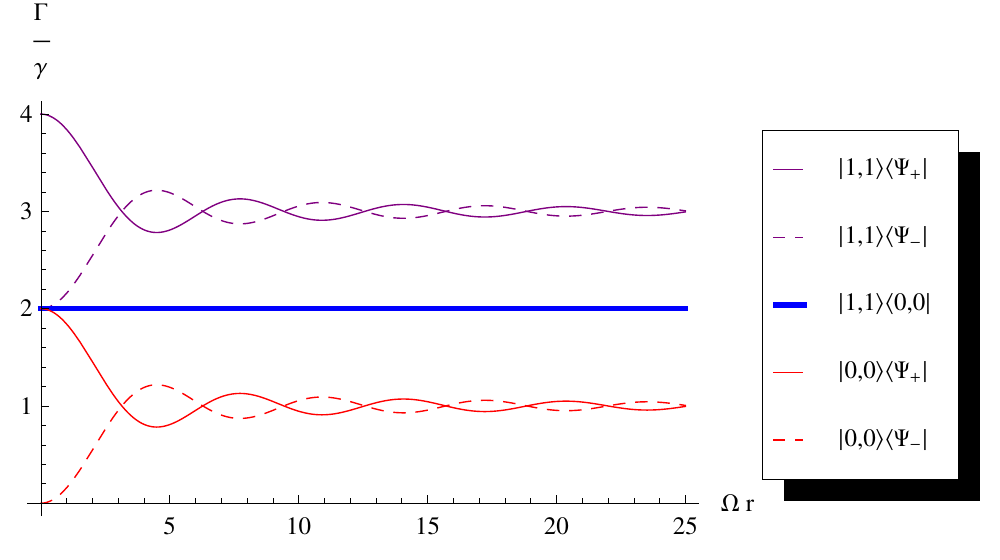}
\caption{Decoherence rates of the (zeroth-order) non-stationary operators for two resonant atoms in a zero-temperature environment at varying separation distance.
The legend indicates the matrix elements they correspond to in the order they occur at the vertical axis.}
\label{fig:Decay1b}
\end{figure}
In Figs.~\ref{fig:Decay1a}--\ref{fig:Decay1b} we plot all relaxation rates associated with the two-atom system as a function of proximity,
where $\gamma$ is specifically the decoherence rate of a single isolated atom.
These are quantities that can be calculated from the RWA-Lindblad equation, and our results are consistent with those reported in \cite{Ficek06}.
For large separation the decay rates for $\ket{\Psi_{\!\pm}} \equiv \left( \ket{0,\!1} \pm \ket{1,\!0} \right)/\sqrt{2} $ are $1+1$ times $\gamma$
(which would be $ N\gamma $ for $ N $ atoms), as the noise processes are independent and the decay rates are additive.
Whereas at proximity they become $0$ and $N^2$ times $\gamma$ for $ \ket{\Psi_{\!-}} $ and $ \ket{\Psi_{\!+}} $ respectively, as the noise processes are maximally correlated and display destructive and constructive interference.

\begin{figure}
\includegraphics[width=0.4\textwidth]{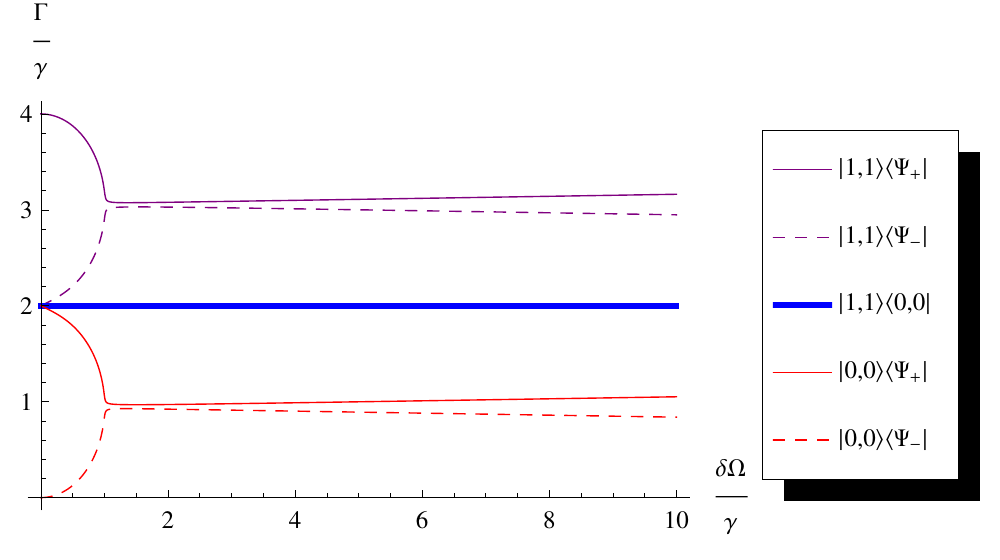}
\caption{Decoherence rates of the (zeroth-order) non-stationary operators for two atoms in a zero-temperature environment at varying detuning and vanishing separation, $ r_{12} \ll \Omega_1,\Omega_2 $, with $\gamma = \langle \Omega \rangle /100$.
The legend indicates the matrix elements they approximately correspond to for small detunings (in the order they occur at the vertical axis) as to compare with Fig.~\ref{fig:Decay1b}.}
\label{fig:Decay2}
\end{figure}
In Fig.~\ref{fig:Decay2} we plot all non-stationary decoherence rates associated with the two-atom system as a function of detuning.
These are new results that cannot be determined from an RWA-Lindblad equation, as for slight detuning there are near-stationary contributions to the dynamics (in the interaction picture) which cannot be discarded.
To achieve a dark state, the tuning of the two atoms must be much better than the dissipation, $\delta \Omega \ll \gamma$,
which counter-intuitively implies that weak-dissipation is not always desirable to preserve coherence.
However, this condition makes more sense if thought of in another way: The dark state arises from the destructive interference of the emission from the two atoms.
If the emission from each atom is characterized by center frequency $ \Omega_n $ and an emission line width $ \gamma $, then the condition $\delta \Omega \ll \gamma$ simply specifies that the emission lines of the atoms must overlap enough that their emissions are not distinguishable from one another.
This allows the required destructive interference.

\subsubsection{The Atomic Seesaw} \label{sec:Seesaw}
One behavior which is qualitatively different from the closed-system evolution is the damped oscillations between the singly-excited states.
More specifically for any initial state of the form
\begin{align}
\boldsymbol{\rho}_0 &= \begin{array}{l} \ket{0,\!1} \\ \ket{1,\!0} \end{array} \left[ \begin{array}{cc} a+\delta & +\imath \, b \\ -\imath \, b & a-\delta \end{array} \right] \begin{array}{l} \bra{0,\!1} \\ \bra{1,\!0} \end{array} \, ,
\end{align}
with all positive coefficients,
then in addition to the Bell-state decay one will also have damped oscillations of the form
\begin{align}
& \left[ \delta \cos(f_1 t) + b \sin(f_1 t) \right] e^{-\gamma_1 t} \left( \ket{0,\!1}\!\!\bra{0,\!1} - \ket{1,\!0}\!\!\bra{1,\!0} \right) \nonumber \\
+ \imath & \left[ b \cos(f_1 t) - \delta \sin(f_1 t) \right] e^{-\gamma_1 t} \left( \ket{0,\!1}\!\!\bra{1,\!0} - \ket{1,\!0}\!\!\bra{0,\!1} \right)
\end{align}
which can oscillate from one excited state to the other excited state with the frequency
\begin{eqnarray}
f_1 &=& 2 \, \tilde{\gamma}_0 \frac{\cos(\Omega r)}{r} \, ,
\end{eqnarray}
for all temperatures.
As this oscillation rate becomes arbitrarily large with proximity, it should be physically observable in some systems.
Though of course, this model does break down for sufficiently small separations.

\subsection{The dark states}\label{sec:DarkState}
First we will review the normal conditions for which dark and bright (sub and super radiant) states emerge for two atoms in this model.
Then in Sec.~\ref{sec:Natom} we will discuss how at finite temperature, and without a RWA-interaction Hamiltonian, two classes of (Dickie) dark states emerge for $N$ atoms, which we designate proper and improper.

All stationary (and thus decoherence-free) states $\boldsymbol{\rho}_\mathrm{D}$ of the open-system must satisfy the relation
\begin{eqnarray}
\boldsymbol{\mathcal{L}} \, \boldsymbol{\rho}_\mathrm{D} &=& 0 \, ,
\end{eqnarray}
and are thus right eigen-supervectors of the Liouvillian with eigenvalue $0$.
As the Liouvillian is not Hermitian, there is no trivial correspondence between the left and right eigen-supervectors.
The super-adjoint of the master equation \cite{Breuer02} time-evolves system observables
and for closed systems can be contrasted
\begin{eqnarray}
\boldsymbol{\mathcal{L}}_0 \, \boldsymbol{\rho} &=& -\imath [ \mathbf{H} , \boldsymbol{\rho} ] \, , \\
\boldsymbol{\mathcal{L}}_0^\dagger \, \mathbf{S} &=& +\imath [ \mathbf{H} , \mathbf{S} ] \, .
\end{eqnarray}
The left eigen-supervector $\mathbf{S}_\mathrm{D}^\dagger$ corresponding to  $\boldsymbol{\rho}_\mathrm{D}$ must therefore satisfy
\begin{eqnarray}
\boldsymbol{\mathcal{L}}^\dagger \, \mathbf{S}_\mathrm{D} &=& 0 \, .
\end{eqnarray}
So for every stationary or decoherence-free state $\boldsymbol{\rho}_\mathrm{D}$ there is a symmetry operator $\mathbf{S}_\mathrm{D}$
whose expectation value is a constant of the motion.
The thermal state or reduced thermal state is such a state.
In the limit of vanishing coupling strength, this state is the familiar Boltzmann thermal state.
One can check that the symmetry operator in this case is proportional to the identity and corresponds to $\mathrm{Tr}[\boldsymbol{\rho}]$ being a constant of the motion.

For two resonant dipoles, with  $\Omega_n = \Omega$, there is another stationary state in the limit of vanishing separation $r_{12}=r$.
Because of degeneracy, any superposition of states
\begin{eqnarray}
\ket{\Psi} &=& a_1 \ket{1,\!0} + a_2 \ket{0,\!1} \, ,
\end{eqnarray}
is also an energy state and therefore annihilated by both $\boldsymbol{\mathcal{L}}_0$ and $\boldsymbol{\mathcal{L}}_0^\dagger$.
Further note that for vanishing separation, the field processes associated with $\mathbf{A}(\mathbf{r},t)$ become exactly correlated and identical.
Their contributions to the interaction Hamiltonian can then be collected into
\begin{equation}
\mathbf{H}_{\mathrm{I}_1} + \mathbf{H}_{\mathrm{I}_2} \,=\, \left( \boldsymbol{\sigma}_{\!x_1} + \boldsymbol{\sigma}_{\!x_2} \right) \mathbf{d} \cdot \mathbf{A}(\mathbf{r}) \,=\, \boldsymbol{\Sigma}_{x} \, \mathbf{d} \cdot \mathbf{A}(\mathbf{r}) \, .
\end{equation}
Next we note the equality
\begin{eqnarray}
\boldsymbol{\Sigma}_{x}  \ket{1,\!0} &=& \boldsymbol{\Sigma}_{x}  \ket{0,\!1} \, ,
\end{eqnarray}
so that for the Bell states
\begin{eqnarray}
\ket{\Psi_{\!\pm}} &\equiv& \frac{1}{\sqrt{2}} \left\{ \ket{1,\!0} \pm \ket{0,\!1} \right\} \, ,
\end{eqnarray}
the noise adds destructively for $\ket{\Psi_{\!-}}$ and constructively for $\ket{\Psi_{\!+}}$.
Therefore $\ket{\Psi_{\!-}}$ is a decoherence-free state (dark state) of the open system for vanishing separation and at resonance, regardless of coupling strength or temperature.
And whereas $\ket{\Psi_{\!-}}$ appears dark (sub-radiant), $\ket{\Psi_{\!+}}$ appears bright (super-radiant).
[Note that for anti-parallel dipoles, these roles will be reversed due to the anti-correlated noise.]

In this particular case the left and right eigen-supervectors are equivalent,
and so it is the dark-state component $\bra{\Psi_{\!-}} \boldsymbol{\rho} \ket{\Psi_{\!-}}$ which is a constant of the motion.
However, unlike the thermal state, if the separation is no longer vanishing then this is not some perturbative limit of a stationary state but of a very long-lived state.
The final constant of motion, which we have validated by analyzing the eigen-system of $\boldsymbol{\mathcal{L}}$, corresponds to the coherence between the ground state and the dark state or $\bra{0,\!0}\boldsymbol{\rho}\ket{\Psi_{\!-}}$.
Using these constants of motion, for two very close dipoles in a zero-temperature environment with initial state $\boldsymbol{\rho}_0$,
the system will relax into the state
\begin{align}
\boldsymbol{\rho}_1 =&\; \left( 1-b \right) \ket{0,\!0}\!\!\bra{0,\!0} + b \ket{\Psi_{\!-}}\!\!\bra{\Psi_{\!-}} \label{eq:DarkRelax} \\
& + c \ket{0,\!0}\!\!\bra{\Psi_{\!-}} + c^* \ket{\Psi_{\!-}}\!\!\bra{0,\!0} \, ,  \nonumber \\
b \equiv&\; \bra{\Psi_{\!-}}\boldsymbol{\rho}_0\ket{\Psi_{\!-}} \, , \\
c \equiv&\; \bra{0,\!0}\boldsymbol{\rho}_0\ket{\Psi_{\!-}} \, ,
\end{align}
to zeroth order in the system-environment coupling,
whereupon the system has bipartite entanglement $b$.

While our (regularized) model is well behaved in the mathematical limit $ r \rightarrow 0 $, it is important to remember
that physically the model is no longer valid for sufficiently small $r$.
At small enough $r$ other terms would come into play, including electrostatic interaction, and eventually the atoms would
cease to even be distinct.  We are assuming that this scale is much smaller than all other scales in our model (except perhaps the cutoff).
This means that we can sensibly consider cases where $r$ is small compared to the other parameters, but $r$ cannot vanish completely.

Since the coefficients of our master equation are continuous in $r$,
it is useful to consider $r=0$ to understand the limiting behavior as $r$ becomes small.  The existence of the dark state we've discussed
at $r=0$ means that this state will be almost completely dark when $r$ is small; thus, any initial state $\boldsymbol{\rho}_0$ will first relax approximately into the state given in Eq.~\eqref{eq:DarkRelax} within the ordinary relaxation timescale $\gamma$,
and then on a much longer relaxation timescale $\tau$, where roughly $1/\tau \approx \gamma (\Omega r)^2$  for small $r$,
the system will fully thermalize.
However, this expression for the dark state is only to zeroth-order in the system-environment coupling.
In order to understand the subsequent final state of decay one needs the second-order asymptotics that we discuss in Sec.~\ref{sec:asymptotic}.

Finally we would note that this ``dark state'' is a very general feature of resonant multipartite systems with similar linear couplings to a shared environment.
One can rather easily work out that for a pair of resonant linear oscillators with these same noise correlations
the sum mode is thermalized, and the difference mode is decoherence free for vanishing separation.
The separation dependence of the entanglement dynamics of two resonant oscillators was considered in Ref.~\cite{Lin09}, while that of  (effectively) two very close oscillators was considered in Ref.~\cite{Paz08,Paz09}.

\subsubsection{Proper and improper dark states for $N$-atoms}\label{sec:Natom}
Here we specifically consider the finite-temperature theory of Dickie subradiance wherein we neglect all forces between the atoms,
including the $1/r^3$ electrostatic (dipole-dipole) and $1/r$ magnetostatic (orbital-orbital) interactions.
It should be noted that for $N > 2$ the inclusion of atomic forces does make for significant differences in the theory which we do not consider here \cite{Friedberg74,Coffey78}.

\begin{figure}
\includegraphics[width=0.4\textwidth]{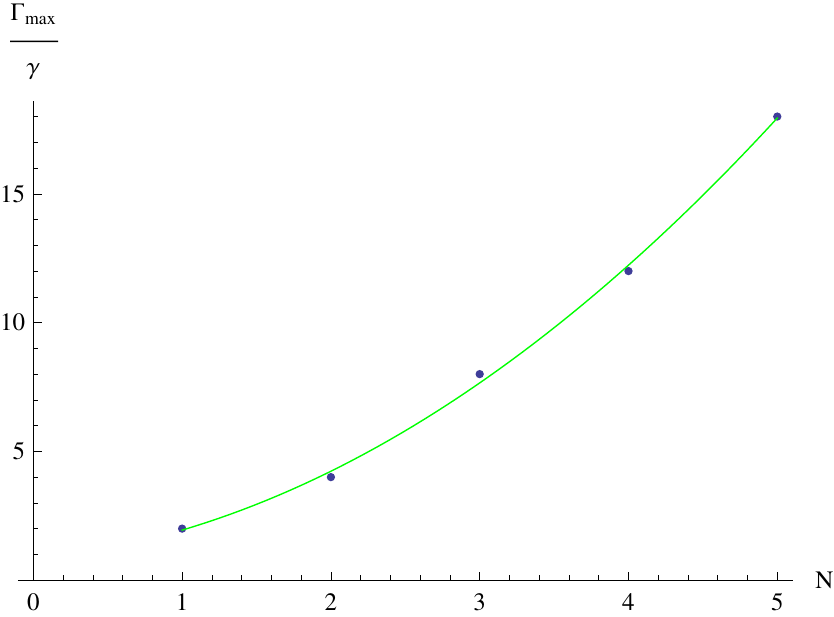}
\caption{Maximal (over states) second-order decay rates as a function of the number of atoms $N$, at zero temperature and in close proximity.
The solid curve denotes the best quadratic fit and has a corresponding p-value of $2.4\%$, which is fairly significant in corroborating an $N^2$ dependence.
}
\label{fig:Super}
\end{figure}

The sub-radiant dark state achieves destructive interference in the environmental noise (and thus little-to-no emission) while the bright state achieves constructive interference in the noise (and thus near-maximal emission).
For the super-radiant bright state one essentially couples the system to $N$ copies of the same field process associated with $\mathbf{A}(\mathbf{r},t)$ and therefore the super-radiant emission rate \emph{can} be proportional to $N^2$.
An $N^2$ dependence does appear the case as we demonstrate in Fig.~\ref{fig:Super}.
The emission rate is (perturbatively) determined by the noise correlation (the square of the noise process).
Both results differ having from $N$ independent noise processes where one can simply add the $N$ independent noise correlations which results in an emission rate at most proportional to $N$.
Up to this point, the physics in this subsection is all well known.

Following the previous approach in this section, we define a \emph{proper dark state} as an atomic state annihilated by $\boldsymbol{\mathcal{L}}_0$ and $\mathbf{H}_\mathrm{I}$ regardless of the state of the environment.
Let us consider an assembly of $N$ resonant dipoles at close proximity.
We first note that the superposition
\begin{align}
\ket{\Psi} &= \sum_{\sum s_n = S} a_{s_1,s_2,\cdots,s_N} \ket{s_1,s_2,\cdots,s_N} \, ,
\end{align}
of energy states with the same total excitement $S$ is also an energy state and therefore annihilated by $\boldsymbol{\mathcal{L}}_0$.
Defining the collective spin operator
\begin{eqnarray}
\boldsymbol{\Sigma}_x &=& \sum_n \boldsymbol{\sigma}_{\!x_n} \, , \label{eq:SigX}
\end{eqnarray}
such that the interaction Hamiltonian can be expressed
\begin{eqnarray}
\mathbf{H}_\mathrm{I} &=& \boldsymbol{\Sigma}_x \, \mathbf{d} \cdot \mathbf{A}(\mathbf{r}) \, ;
\end{eqnarray}
a proper dark state must then satisfy $\boldsymbol{\Sigma}_x \ket{\Psi_{-}} = 0$ and will be decoherence free.
For $N=2$ this is the familiar Bell state that we've already labeled $ \ket{\Psi_{-}} $.

In considering large $ N $ the structure is essentially just what was studied by Dicke \cite{Dicke54}, so following that approach we define collective $ y $  and $ z $ spin operators $ \boldsymbol{\Sigma}_y $ and $ \boldsymbol{\Sigma}_z $ as well as raising and lowering operators $ \boldsymbol{\Sigma}_+ $ and $ \boldsymbol{\Sigma}_- $, analogously to Eq.~\eqref{eq:SigX}, as well as
$ \boldsymbol{\Sigma}^2 = \boldsymbol{\Sigma}_x^2 + \boldsymbol{\Sigma}_y^2 + \boldsymbol{\Sigma}_z^2$.
And we can note that the free Hamiltonian for the atoms only differs from $ \boldsymbol{\Sigma}_z $ by a multiple of the identity, so all the eigenstates of that Hamiltonian are also eigenstates of $ \boldsymbol{\Sigma}_z $.
A basis for the Hilbert space of the system can be specified by  the eigenstates of $ \boldsymbol{\Sigma}^2 $ with eigenvalues $j(j+1) $ and $ \boldsymbol{\Sigma}_z $ with eigenvalues $ m $ (though for $ N>2 $ there will be degeneracy, so that additional quantum numbers are needed to identify a specific state).
The dark state we seek must then satisfy $ \boldsymbol{\Sigma}_z  \ket{\Psi_{-}} = m  \ket{\Psi_{-}} $ and $\boldsymbol{\Sigma}_x \ket{\Psi_{-}} = 0$.  As the discussion in \cite{Dicke54} implies, only states with $ j=0 $ and $ m=0 $ can satisfy these requirements simultaneously.
Such states only occur when $ N $ is even, and that set of states has dimension $ N! / \left[ (N/2+1)! (N/2)! \right] $.
These are also the dark states in the RWA, as they are in the null space of both $ \boldsymbol{\Sigma}_+ $ and $ \boldsymbol{\Sigma}_- $.
For $N=4$ these states take the form
\begin{align}
\ket{\Psi_-} =&\; a_1 (\ket{0,\!0,\!1,\!1}+\ket{1,\!1,\!0,\!0}) + a_2 (\ket{0,\!1,\!0,\!1}+\ket{1,\!0,\!1,\!0}) \nonumber \\
& + a_3 (\ket{0,\!1,\!1,\!0}+\ket{1,\!0,\!0,\!1}) \, , \\
0 =&\; \sum_n a_n \, ,
\end{align}
where every pair in parenthesis is spin-flip symmetric.
One can easily check that any such state is annihilated by $\boldsymbol{\Sigma}_x$.

However, more generally we define an \emph{improper dark state} as one only annihilated by $\boldsymbol{\mathcal{L}}$ and not $\mathbf{H}_\mathrm{I}$
(i.e., stationary in the coarse-grained open-system dynamics but not in the full closed system dynamics), thus being dependent upon the state of the environment and even the coupling strength.
In the simplest case we can consider the zero-temperature environment.
For the second-order dynamics, upward transitions are automatically ruled out from the lack of thermal activation.
Rather than investigating the master equation, for zero temperature we can then simply demand that the lowest-order decay transitions are vanishing,
meaning that if $\ket{\Psi_-^S}$ has total excitation $S$, then $\bra{S'} \boldsymbol{\Sigma}_x \ket{\Psi_-^S} = 0$ for all $S' \leq S$ lesser and equally excited states.
We  can also state this in terms of the collective spin operators we have defined, by saying that we demand that $\ket{\Psi_-}$ is an eigenstate of $ \boldsymbol{\Sigma}_z $ with eigenvalue $ m $, and that all matrix elements onto states with lower $ m^{\prime} $ values must vanish.
Since $ \boldsymbol{\Sigma}_x = \frac{1}{2} \left( \boldsymbol{\Sigma}_+ + \boldsymbol{\Sigma}_- \right) $, we know that there will be non-vanishing matrix elements onto states with $ m^{\prime} = m - 1 $ unless $ m = -j$.
So any state with $ m = -j$ is an improper dark state at zero temperature, and there are $ N! (2j+1)/ \left[ (N/2+j+1)! (N/2-j)! \right] $ such states \cite{Dicke54}.
Interestingly, for the RWA-interaction Hamiltonian such states (when combined with a vacuum field) are also stationary states but of the closed-system dynamics.
For $N=3$ and at zero temperature, all such dark states can be expressed
\begin{align}
\ket{\Psi_-} &= a_1 \ket{1,\!0,\!0} + a_2 \ket{0,\!1,\!0} + a_3 \ket{0,\!0,\!1} \, , \\
0 &= \sum_n a_n \, ,
\end{align}
for weak coupling to the field.
Numerical investigation of the second-order master equation, and counting the number of null eigen-operators of $\boldsymbol{\mathcal{L}}$ for different $N$ and $T$,
reveals that these dark states also exist for positive temperature, but they take on a different, temperature-dependent form.
Though it is clear how the simple criteria leading up to resolution of the zero-temperature improper dark states fails to carry over into the finite-temperature regime,
it is not clear how these finite-temperature improper dark states can be calculated without the aid of a finite-temperature master equation,
nor what simpler properties they might exhibit.
Moreover, they may only exist perturbatively in the second-order master equation: the fourth-order master equation might assign them fourth-order emission rates.

\section{Second-order asymptotic solution}\label{sec:asymptotic}
\subsection{Canonical perturbation theory}
In this section we specifically address the asymptotic state $\boldsymbol{\rho}(\infty)$ to second order.
As we have previously mentioned, no second-order master equation can fully determine all second-order state information,
including measures of entanglement.
However, as we will point out, the second-order master equation is sufficient for determining the asymptotic state at high temperature.
Moreover, at zero temperature one can apply canonical perturbation theory to the closed system + environment Schr\"{o}dinger equation and bypass the master equation formalism entirely.

To zeroth order in the system-environment interaction, the asymptotic steady state is Boltzmann, which can be expressed
\begin{eqnarray}
\boldsymbol{\rho}_{T} &=& \prod_n \boldsymbol{\rho}_{T_n} \, , \label{eq:ZOAsymptotic} \\
\boldsymbol{\rho}_{T_n} &\equiv& \frac{1}{2} \left[ 1 - \tanh\!\left( \frac{\Omega_n}{2T} \right) \boldsymbol{\sigma}_{\!z_n} \right] \, ,
\end{eqnarray}
in terms of Pauli matrices.
The asymptotic state of the second-order master equation is consistent with this result and can additionally provide some of the second-order corrections $\boldsymbol{\delta\! \rho}_T$ via the constraint
\begin{eqnarray}
\boldsymbol{\mathcal{L}}_0 \{ \boldsymbol{\delta\! \rho}_T \} + \boldsymbol{\mathcal{L}}_2 \{ \boldsymbol{\rho}_{T} \} &=& 0 \, . \label{eq:SS2ndO}
\end{eqnarray}
These will specifically be the off-diagonal or non-stationary perturbations.
In general, to find the second-order corrections to the diagonal elements of the density matrix one needs to compute contributions from the fourth-order Liouvillian \cite{Accuracy}.

It has been shown \cite{Mori08,QOS} that for non-vanishing interaction with the environment the off-diagonal elements of the asymptotic state match the reduced thermal state
\begin{equation}
\boldsymbol{\rho}_{\beta} \equiv \frac{1}{Z_\mathrm{C}(\beta)} \tr_E \left[ e^{-\beta( \mathbf{H} + \mathbf{H}_\mathrm{E} + \mathbf{H}_\mathrm{I} )} \right] \, ,
\end{equation}
where $Z_\mathrm{C}(\beta)$ is the partition function of the system and environment with non-vanishing interaction.
We will refer to $\boldsymbol{\rho}_{\beta}$ as the thermal Green's function; this function can be expanded perturbatively in the system-environment coupling as
\begin{eqnarray}
\boldsymbol{\rho}_{\beta} &=& \frac{1}{Z_0(\beta)} e^{-\beta \, \mathbf{H}} + \boldsymbol{\delta\!\rho}_{\!\beta} + \cdots \, , \label{eq:Gseries}
\end{eqnarray}
where $Z_0(\beta)$ is the partition function of the free system.
The second-order corrections are given by
\begin{equation}
\bra{\omega_i} \boldsymbol{\delta\!\rho}_{\!\beta} \ket{\omega_j} = \sum_{nmk} \frac{R_{ijk}^{nm}}{Z_0(\beta)}  \bra{\omega_i} \boldsymbol{\sigma}_{\!x_m} \ket{\omega_k} \bra{\omega_k} \boldsymbol{\sigma}_{\!x_n} \ket{\omega_j} \, . \label{eq:G(beta)}
\end{equation}
All terms with $\omega_i = \omega_j$ are zero, so that this expression gives no correction to the diagonal elements of the density matrix.
Otherwise, the (non-resonant) off-diagonal coefficients are given by
\begin{align}
& \left. R_{ijk}^{nm} \right|_{\omega_i \ne \omega_j} \equiv  \mathrm{Im}\!\left[ e^{-\beta \omega_k} \frac{\mathcal{A}_{nm}(\omega_{ik})\!-\!\mathcal{A}_{nm}(\omega_{jk})}{\omega_i - \omega_j}  \right] \nonumber \\
& + \mathrm{Im}\!\left[ \frac{e^{-\beta \omega_i} \mathcal{A}_{mn}(\omega_{ki}) \!-\! e^{-\beta \omega_j} \mathcal{A}_{mn}(\omega_{kj})}{\omega_i-\omega_j} \right] \, , \label{eq:Rij}
\end{align}
with the free ground-state energy set to zero.
These coefficients agree perturbatively with those from Eq.~\eqref{eq:SS2ndO}.
Because such an expansion is inherently secular in $\beta$, it is valid only at a sufficiently high temperature such that the perturbations are small compared to the smallest Boltzmann weight,
\begin{equation}
\frac{\gamma}{\Omega} \ll e^{-\beta ( \Omega_n + \Omega_m )} =
\left( \frac{\bar{n}(\Omega_n,T)}{\bar{n}(\Omega_n,T)\!+\!1} \right) \left( \frac{\bar{n}(\Omega_m,T)}{\bar{n}(\Omega_m,T)\!+\!1} \right) \label{eq:HighTCond}.
\end{equation}
The expansion does not apply at lower temperatures.
Reliability of the expansion at higher temperature suggests that the diagonal corrections to the asymptotic state must be suppressed there.

Since neither the second-order master equation nor the perturbative expansion of the thermal Green's function can give the full low-temperature solution, including diagonal corrections, it appears that in general this will require the fourth-order master equation coefficients. However, at zero temperature the thermal state is simply the ground state of the total system-environment Hamiltonian.
This ground state can be calculated perturbatively from the Hamiltonian as usual in a closed system, and the zero-temperature reduced thermal state follows directly.  All three of these formalisms are fully consistent as shown in Ref.~\cite{QOS}.  At zero temperature the off-diagonal second-order corrections to the asymptotic state are still of the form given in Eqs.~\eqref{eq:G(beta)} and \eqref{eq:Rij}, with the coefficients evaluated in the limit $ \beta \rightarrow \infty $.
The diagonal (and resonant) perturbations are given by
\begin{align}
& \lim_{\beta \to \infty} \left. R_{ijk}^{nm} \right|_{\omega_i = \omega_j} =  \label{eq:Cinterpolate} \\
& \lim_{\beta \to \infty} \mathrm{Im}\!\left[ e^{-\beta \omega_k} \frac{d}{d\omega_i} \mathcal{A}_{nm}(\omega_{ik}) + e^{-\beta \omega_i} \frac{d}{d\omega_i} \mathcal{A}_{mn}(\omega_{ki}) \right] \, , \nonumber
\end{align}
where only a handful of terms are non-vanishing.
We note that the expression inside the limit in Eq.~\eqref{eq:Cinterpolate} has both the correct low and high-temperature limits, so it may be roughly correct for all temperatures, but we have yet to fully investigate the fourth-order master equation.

For most regimes the second-order thermal state can now be expressed entirely in terms of the second-order master equation coefficients and limits thereof.
Therefore we can say that the environmentally induced correlations do vanish for large separations with a power-law decay like $1/r$ and $1/r^2$.

\subsection{Entanglement of Two Atoms}
Now we will consider the bipartite entanglement between any two atoms, labeled $n$ and $m$ in a common quantum field. We begin with some remarks that apply to any system of two qubits.  We focus on the late-time dynamics of this system; we will compute the reduced density matrix for their asymptotic state $\boldsymbol{\rho}_{nm}$ and derive the asymptotic value of entanglement between these two atoms. We will see that this computation will also allow us to show that all entangled initial states become disentangled at a finite time.

To quantify the bipartite entanglement we will use Wootters' concurrence function \cite{Wootters98}, which is a monotone with a one-to-one relationship to the \emph{entanglement of formation} for two qubits.  The concurrence is defined as
\begin{align}
C \! \left( \boldsymbol{\rho}_{nm} \right)  =& \max \left\{ 0,\unc \! \left( \boldsymbol{\rho}_{nm} \right) \right\} \\
\unc\! \left( \boldsymbol{\rho}_{nm} \right) =& \sqrt{\lambda_{1}}-\sqrt{\lambda_{2}}-\sqrt{\lambda_{3}}-\sqrt{\lambda_{4}}
\end{align}
where $\lambda_{1} \ge \lambda_{2} \ge \lambda_{3} \ge \lambda_{4}$ are the eigenvalues of the matrix
\begin{equation}
\boldsymbol{\rho}_{nm} \, \tilde{\boldsymbol{\rho}}_{nm} \equiv \boldsymbol{\rho}_{nm} \left( \boldsymbol{\sigma}_{\!y_n} \, \boldsymbol{\sigma}_{\!y_m} \, \boldsymbol{\rho}_{nm}^{*} \, \boldsymbol{\sigma}_{\!y_n} \, \boldsymbol{\sigma}_{\!y_m}  \right) \, ,
\end{equation}
which are always non-negative.
A two-qubit state is entangled if and only if $ \unc > 0 $.
It is important to note that $ \unc \left( \boldsymbol{\rho} \right)$ is a continuous function of the matrix elements of $ \boldsymbol{\rho} $ (since the eigenvalues of a matrix can be written as a continuous function of the matrix elements \cite{Stewart01});
this then implies that any density matrix with $ \unc < 0 $ lies in the interior of the set of separable states (with every sufficiently nearby state also separable), while states with $ \unc > 0 $ lie in the interior of the set of entangled states.
States with $ \unc = 0 $ are separable but include states that lie on the boundary between the two sets, infinitesimally close to both entangled states and the interior of the separable states.
Any separable pure state lies on this boundary \cite{Horodecki09}.

Given the late-time asymptotic state of two atoms $ \boldsymbol{\rho}_{nm} $, one can easily compute the asymptotic entanglement from $ \unc \left( \boldsymbol{\rho}_{nm} \right)$.
Based on the preceding paragraph, however, we know that this will also tell us something qualitatively about the late-time entanglement dynamics.
If $ \unc \left( \boldsymbol{\rho}_{nm} \right) < 0$ then (assuming only continuous evolution in state space) every initial state must become separable at some finite time as it crosses into the set of separable states.
Likewise, if $ \unc \left( \boldsymbol{\rho}_{nm} \right) > 0$ then all initial states lead to entanglement at sufficiently late time and any sudden death of entanglement must be followed by revival.
In models such as ours which have a unique asymptotic state, it is only when $ \unc \left( \boldsymbol{\rho}_{nm} \right) = 0$ that this qualitative feature of the late-time behavior will depend on the initial state, with some entangled states remaining separable after some finite time and others becoming distentangled only asymptotically in the limit $ t \rightarrow \infty $ as in \cite{YuEberly04,Ficek06}.
Previous work has pointed out that the late-time entanglement dynamics can be determined by the asymptotic state in this way \cite{YuEberly07a,TerraCunha07}, with Yu and Eberly \cite{YuEberly07} discussing the role of $ \unc $ in predicting sudden death.
In Refs.~\cite{YuEberly07,TerraCunha07} the authors consider models with multiple steady states, which introduces additional dependence on initial conditions.

\begin{figure}[h]
\includegraphics[width=0.4\textwidth]{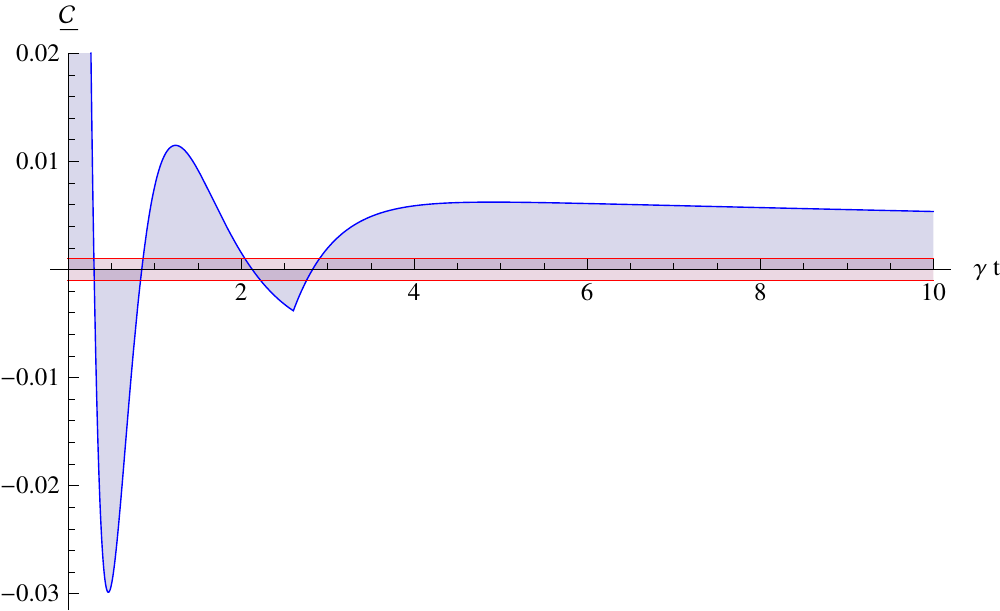}
\caption{\label{fig:EntDynamic}
Unmaximized concurrence for 2 atoms initially in the state $\boldsymbol{\psi}_0=\sqrt{0.1}\ket{0,0} + \sqrt{0.9}\ket{1,1}$ with $\gamma = \Omega/1000$ and $r= 2\pi/20\Omega$,
showing sudden death, revival, and (seemingly) asymptotic disentanglement as the concurrence trends to zero.
(Compare with \cite{Ficek06} Fig.~2.)
The middle, red region spans  $\pm \gamma/\Omega$ and denotes the order of neighborhood around $\unc = 0$ which cannot be resolved by any second-order master equation.}
\end{figure}
\begin{figure}[h]
\includegraphics[width=0.4\textwidth]{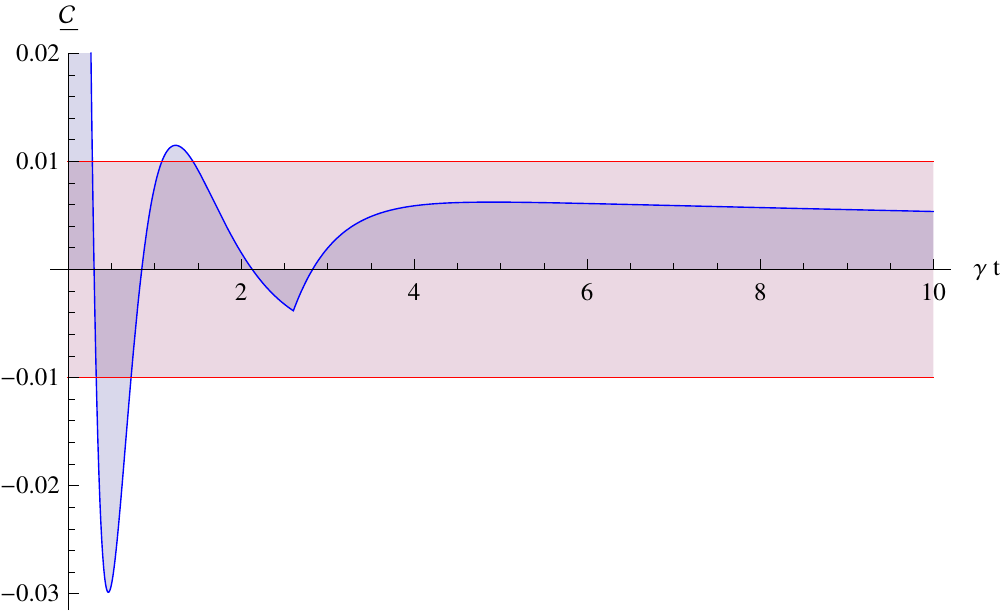}
\caption{\label{fig:EntDynamic2}The same plot as Fig.~\ref{fig:EntDynamic}, but with the larger decay rate $\gamma = \Omega/100$.
In this case the prediction of entanglement revival is swamped by uncertainties inherent in the second-order master equation, the order of which are spanned by the middle, red region.}
\end{figure}
It can be seen that none of the foregoing discussion is specific to the concurrence; it would apply to any quantity that is a continuous function of the density matrix, takes on negative values for some separable states, and is an entanglement monotone when non-negative.  If we have such an unmaximized entanglement function $ \unmax{\gem} $ from which an entanglement monotone can be defined by $ \gem = \max \left\{ 0, \unmax{\gem} \right\} $, then we can use it just as we have discussed using $ \unc $ above.
\begin{figure}[h]
\includegraphics[width=0.4\textwidth]{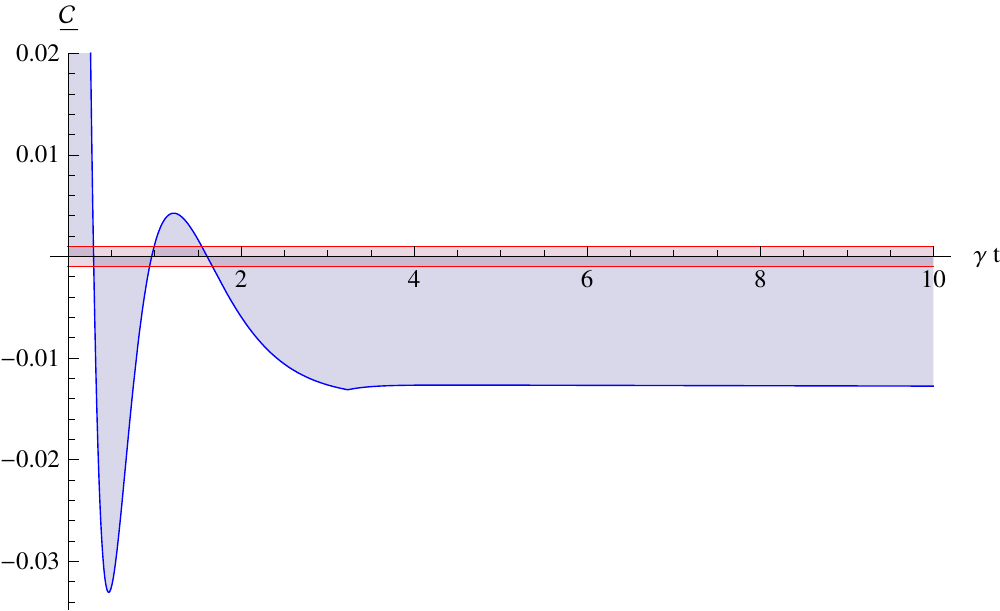}
\caption{\label{fig:EntDynamicT}
The same plot as Fig.~\ref{fig:EntDynamic}, but at finite temperature with $T = \Omega/5$.
In this case sudden death of entanglement (after a brief revival) can be confirmed with the second-order master equation, but only by using an unmaximized entanglement monotone.}
\end{figure}
As illustrated with a specific example in Fig.~\ref{fig:EntDynamicT}, entanglement sudden death occurs because the unmaximized entanglement function asymptotes towards a negative value, whereas any entanglement monotone (derived from $ \unmax{\gem} $ or otherwise) cannot go below zero, leading to an abrupt sudden death of entanglement when $ \unmax{\gem} $ becomes negative.

\begin{figure}[h]
	\centering
	\subfloat[Pure Asymptotic State]{
		\includegraphics[width=0.4\textwidth]{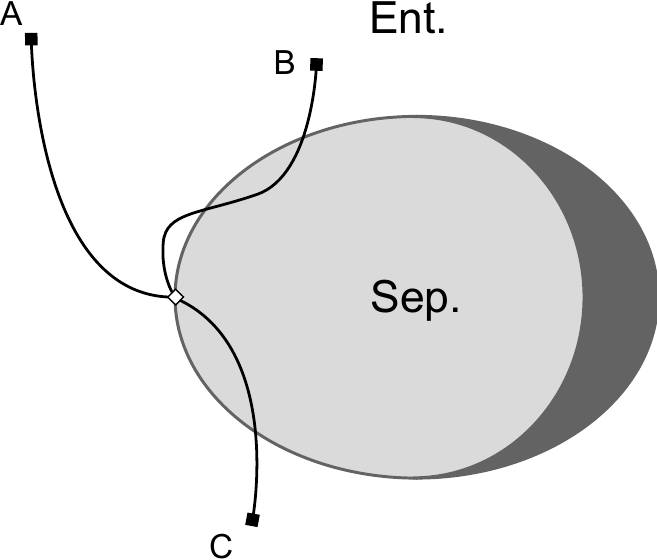}
		\label{fig:purestatespace}
	}\\
	\subfloat[Mixed Asymptotic State]{
		\includegraphics[width=0.4\textwidth]{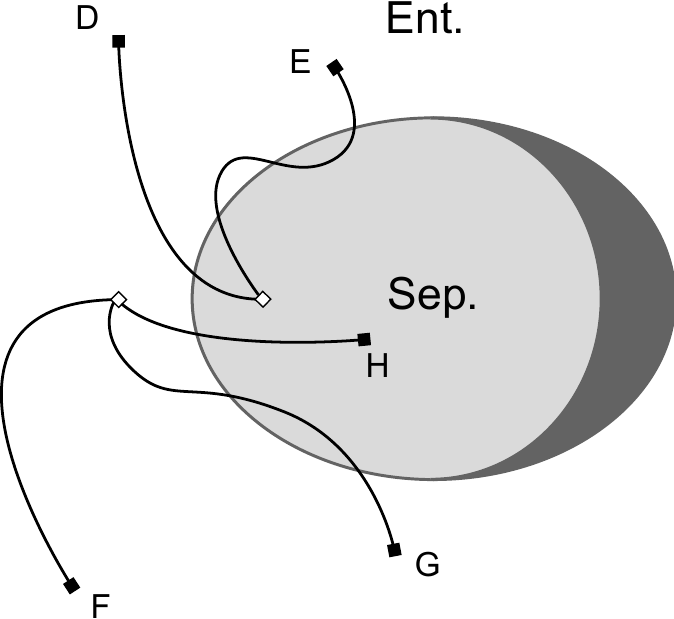}
		\label{fig:mixedstatespace}
	}
	\caption{
		A schematic representation of the evolution in state space.  The white area represents entangled states
		($ \unc > 0 $),
		while the gray areas represent separable states $ \unc  \le 0 $
		with the dark gray representing states with $ \unc = 0 $.
		The asymptotic state is represented by $\diamond$, while initial states are represented by $ {\scriptstyle \blacksquare} $.
		In (a) we have the asymptotic state on the boundary as in the
		zeroth-order at $ T=0 $.  In (b) two scenarios are shown that can arise from
		a small perturbation moving the asymptotic state off the boundary, into the interior of one of the two
		sets.  This illustrates how such a perturbation qualitatively changes the late-time entanglement		
		dynamics.
		}
	\label{fig:statespace}
\end{figure}

An important point arises from the facts we have noted about $ \unc $ and separability:
At sufficiently low temperature the $ \mathcal{O}(\gamma) $ corrections to the asymptotic state are required to calculate the sign of $ \unc \left( \boldsymbol{\rho}_{nm} \right)$ and, therefore, even the qualitative features of late-time entanglement dynamics.
Specific examples are plotted in Fig.~\ref{fig:EntDynamic}-\ref{fig:EntDynamic2}.
At zero temperature, the zeroth-order asymptotic state is simply the ground state of the system (assuming no degeneracy at the ground energy) according to Eq.~\eqref{eq:ZOAsymptotic}.
So the zeroth-order asymptotic state is a pure separable state. This means that it lies on the boundary between the entangled and separable states, and in general some initial states will suffer sudden death while others will not, as depicted in Fig.~\ref{fig:purestatespace}.  But any non-zero perturbation, however small, can lead to asymptotic entanglement or can place the asymptotic state in the interior of the separable states, implying sudden death for all initial conditions.  Fig.~\ref{fig:mixedstatespace} shows each of these situations.
Thus, knowing only the zeroth-order asymptotic state one can make no meaningful prediction about late-time entanglement dynamics, and this will always be the case when using the rotating-wave approximation, because it neglects the second-order corrections to the asymptotic state \cite{RWA}.
This makes calculations such as \cite{Ficek06} incapable of correctly predicting these features.
Moreover, as one can see from Fig.~\ref{fig:EntDynamic2}, in certain regimes errors inherent in the second-order master equation can completely swamp almost all predictions of the entanglement dynamics.
For example, with the initial conditions that give rise to Fig.~\ref{fig:EntDynamic}-\ref{fig:EntDynamic2}, one can see that the decay rate must satisfy $\gamma \ll \Omega/100$ in order to ensure that the prediction of revival, made for that situation in Fig. 2 of \cite{Ficek06}, is accurate despite the presence of errors.
This is a general consequence of \cite{Accuracy}  that applies to any calculation resulting in a concurrence comparable to $\gamma / \Omega$.

At positive temperature the zeroth-order asymptotic state is simply the Boltzmann state $ \boldsymbol{\rho}_{T} $, which lies in the interior of the set of separable states \cite{YuEberly07a}, and
\begin{equation}
\boldsymbol{\rho}_{T} \tilde{\boldsymbol{\rho}}_{T} =  \frac{e^{- \left( \Omega_n + \Omega_m \right) /T }}{Z_0(T)^2} \; \boldsymbol{1} \, ,
\end{equation}
so that $ \unc \left( \boldsymbol{\rho}_{T} \right) = -2 e^{- \left( \Omega_n + \Omega_m \right) /(2T) }/Z_0(T)$.  The $ \mathcal{O}(\gamma) $ corrections to $ \boldsymbol{\rho}_{nm} $ will yield order $ \mathcal{O}(\gamma) $ corrections to  $ \boldsymbol{\rho}_{nm} \tilde{\boldsymbol{\rho}}_{nm} $.
Then simply from the definition of $ \unc $ we know that so long as the temperature is sufficiently high that Eq.~\eqref{eq:HighTCond} is satisfied
the corrections to $ \boldsymbol{\rho}_{nm} $ will cause at most $ \mathcal{O}(\gamma) $ corrections to $ \unc \left( \boldsymbol{\rho}_{nm} \right)$ so that it must remain negative. Consequently, the second-order asymptotic state still lies in the interior of the separable states, and all initial states will suffer entanglement sudden death at sufficiently late times.  For lower temperatures it does not appear that the sign of $ \unc \left( \boldsymbol{\rho}_{nm} \right)$ can be generically predicted, and one must find the late-time asymptotic state for the specific system in question which generally requires terms from the fourth-order master equation.

Returning to the specifics of the particular model examined in this paper, from Eq.~\eqref{eq:G(beta)} it can be readily seen that the atoms are correlated in the asymptotic state at all temperatures,
and from our second-order coefficients these correlations experience power-law decay with separation.  However, we find based on Eqs.~\eqref{eq:Gseries},
\eqref{eq:G(beta)}, and \eqref{eq:Rij} that when the high-temperature expansion is valid (according to Eq.~\eqref{eq:HighTCond}) the asymptotic state has $ \unc \left( \boldsymbol{\rho}_{nm} \right) < 0$.
At zero temperature, Eqs.~\eqref{eq:Rij} and \eqref{eq:Cinterpolate} also give $ \unc \left( \boldsymbol{\rho}_{nm} \right) < 0$.  In both cases the asymptotic state lies in the interior of the separable states,
and all initial states become separable permanently after some finite time.  With this property upheld for zero and high temperatures,
we suspect this to be true at all temperatures, making entanglement sudden death a generic feature which happens in every case in this model.
Of course, as discussed in Sec.~\ref{sec:DarkState}, for closely spaced atoms there can be a dark state, so that entanglement persists over a long timescale before eventually succumbing to sudden death.
It should also be noted that, while this examination of the asymptotic behavior tells us that entanglement always remains zero after some finite time,
we do find $ \mathcal{O}(\gamma^0)$ sudden death and revival of entanglement at earlier times for some initial states (similar to \cite{Ficek06}) as depicted in Fig.~\ref{fig:EntDynamic}-\ref{fig:EntDynamicT}.

\begin{figure*}[htb]
\includegraphics[width=0.9\textwidth]{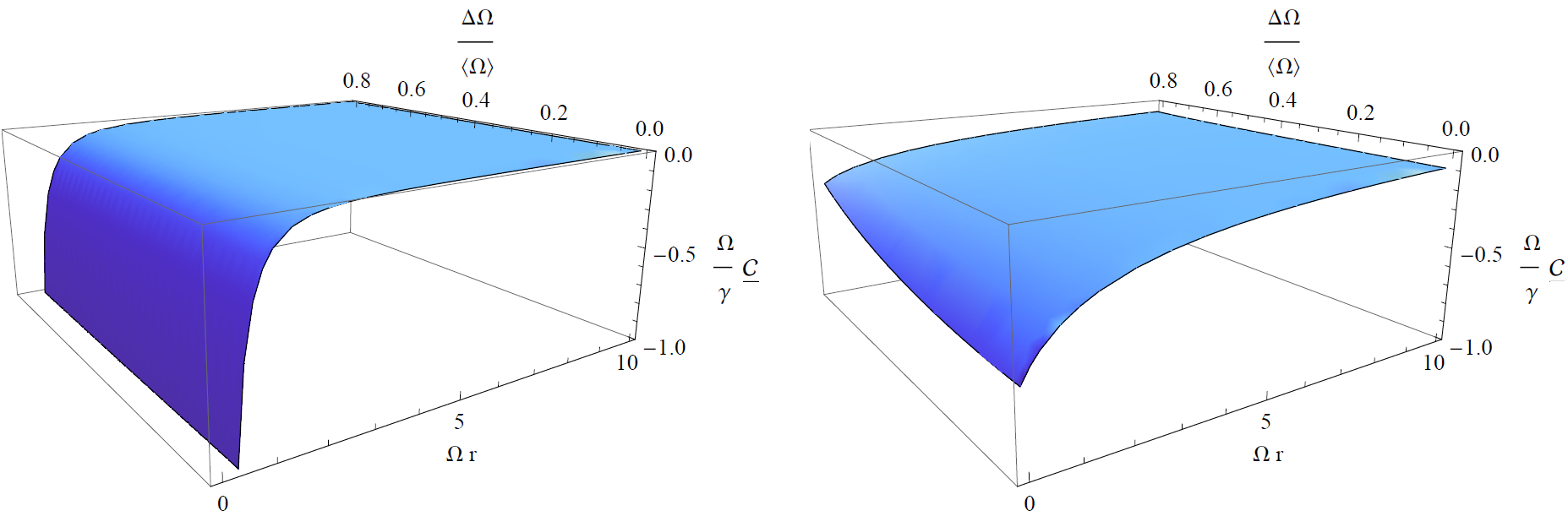}
\caption{Unmaximized concurrence for two resonant atoms at various separation distances and frequency detunings at zero temperature and for $\gamma = \langle \Omega \rangle/100$.
The left plot includes all terms in the model whereas the right plot neglects the $1/r$ magnetostatic interactions, as to inspect the finite contributions.
For the typical RWA-Lindblad equation, these plots would simply be $\unc=0$ given asymptotic relaxation to the free ground state.
The plots here are accurate to second order, which is beyond that of any prediction that can be made with a second-order master equation.}
\label{fig:C}
\end{figure*}
In Fig.~\ref{fig:C} we plot $ \unc $ as it varies with separation distance and frequency detuning.  As a consistency check we also calculated the logarithmic negativity and found it to be consistent with the concurrence to second order.
The behavior of the entanglement is markedly different from that of two oscillators in a field.
The separation dependence of two resonant oscillators was considered in Ref.~\cite{Lin09} and the more general solution will be given in Ref.~\cite{NQBM}.
For two oscillators, there can be asymptotic entanglement if they are held very close and near enough to resonance with each other.
Separation and detuning then causes the entanglement monotones to decay away. For the two-atom case studied here asymptotic entanglement does not exist, and resonant tuning with proximity will only exacerbate the problem.
Permanent sudden death of entanglement occurs because the unmaximized entanglement functions can trend below zero within a finite amount of time and without the need of any asymptotic limit.

\section{Discussion \label{sec:discussion}}

In this paper we have derived the dynamics of a collection of  two-level atoms under a dipole approximation interacting with a common quantized electromagnetic field assuming only weak coupling to the field.
With a careful perturbative analysis we have obtained all dynamical information to second order,
including regimes which cannot be described by RWA-Lindblad equations.
We have also presented a method of finding the zero-temperature asymptotic state to higher accuracy than is possible directly with any second-order master equation, including those derived with the RWA and BMA.
We have used this to show that even at zero temperature the bipartite entanglement between any pair of two-level atoms will undergo sudden death for all initial atomic states, in contrast to the predictions of previous theoretical treatments \cite{Ficek06}. (We will point out specific deficiencies of \cite{ASH06} in a later communication.)
Finally, we have noticed that a class of $N$-atom (Dickie) dark states, which appear normal in the RWA Hamiltonian, are not ordinary dark states and 
are, in fact, temperature dependent.

We have argued that in the RWA there can be inaccuracies in all entries of the density matrix that are of the order of the damping rate $ \gamma $.
By contrast, when represented in the (free) energy basis the solution we have derived here will have off-diagonal elements that are accurate at second-order, having $ \mathcal{O}(\gamma^2) $ errors.
Even in this solution diagonal matrix elements (and matrix elements between any two degenerate energy states) can still have $ \mathcal{O}(\gamma) $ errors, due to a fundamental limitation of any second-order master equation.
However, the expectation of any operator that has vanishing diagonals in the energy basis (including atomic dipole operators), will have only $ \mathcal{O}(\gamma^2) $ errors.  Moreover, unlike some other methods of solution, our solution can be applied when the atoms have distinct (but close) frequencies.

At sufficiently low temperature, the zeroth-order asymptotic state (given by the RWA) is near the boundary between the separable and entangled states, and the small perturbation induced by the environment at $ \mathcal{O}(\gamma) $ can push it into either set.
Depending on which set the perturbed asymptotic states fall into, all states may experience entanglement sudden death or all may become entangled asymptotically.
We have presented a second-order solution for the asymptotic state of any two atoms, which allows us to say decisively that \emph{the zero-temperature asymptotic state of those atoms is separable, and pairwise entanglement of all two-level atoms experiences sudden death regardless of the initial state}.

It should be noted that, for example, in some optical-frequency atomic systems the $ \mathcal{O}(\gamma) $ corrections we discuss can be quite small, with $ \gamma / \Omega $ being perhaps something on the order of $ 10^{-9} $.
Though lowest order corrections to the timescales cannot be ignored (as they are responsible for the presence of dissipation), corrections of this size to the values of the density matrix elements at any instant can easily be considered negligible.
However, in the case of a theoretical study of entanglement sudden death, where one wishes to distinguish asymptotic decay to zero from vanishing in finite time, small perturbations can become vitally important, as they do at low temperature.
And in optical frequency atomic systems at room temperature the thermal-average photon number will be far smaller than $ 10^{-9} $, placing the system deep into what we are considering the low-temperature regime for entanglement dynamics.

We have characterized the sub- and super-radiant states that exist in this model when the RWA is not used.
We have shown that there is still a long-lived, highly-entangled dark state when the atoms have small enough separation, and sudden death of entanglement occurs only on the much longer timescale of decay of this state (assuming it had some population in the initial state).
In this simple model, decoherence-free dark states are achievable for arbitrary temperature and dissipation, whereas typically these factors together are the primary cause of decoherence.
However, for $N>2$, the finite-temperature Dickie dark states fall into two categories: one group is temperature-independent and the other becomes temperature dependent.
This temperature dependence is not seen after employing the RWA.

We close with a few remarks:  1) With the knowledge of distance dependence, to preserve entanglement in time one should place the atoms very close to each other in the field, so as to produce strong cross correlations in the noise.
But at some proximity one must also consider further atom-atom interactions self-consistently within the confines of field theory. 2) Qualitative differences between systems under the two-level and dipole-interaction approximations and harmonic systems suggests a degree of model dependence in some of the phenomena considered; this merits further investigation into the consequences of these approximations.
3) Many other sorts of level structures are relevant to experimental systems, both in terms of the number of levels involved and the angular momentum exchange with the field. The methodology and conceptions developed in this work can be applied for the analysis of the non-Markovian dynamics of more general systems, from which one can perhaps better understand how model-dependent the entanglement behavior considered herewith is.

\section*{Acknowledgment}  This work is supported partially by
NSF Grants PHY-0426696, PHY-0801368, DARPA grant
DARPAHR0011-09-1-0008 and the Laboratory of Physical Sciences.

\bibliography{bib}{}
\bibliographystyle{apsrev4-1}

\end{document}